\documentclass[apj]{emulateapj}
\usepackage{graphics}
\usepackage{alltt}
\usepackage{verbatim}

\def\spose#1{\hbox to 0pt{#1\hss}}
\def\simlt{\mathrel{\spose{\lower 3pt\hbox{$\mathchar"218$}}
    \raise 2.0pt\hbox{$\mathchar"13C$}}}
\def\simgt{\mathrel{\spose{\lower 3pt\hbox{$\mathchar"218$}}
    \raise 2.0pt\hbox{$\mathchar"13E$}}}

\newcommand{\oiii}{\mbox{[\ion{O}{3}]} $\,$}
\newcommand{\oiiiw}{\mbox{[\ion{O}{3}] $\lambda$5007} $\,$}
\newcommand{\oiiiwn}{\mbox{[\ion{O}{3}] $\lambda$5007}}

\newcommand{\ha}{\mbox{H$\alpha$} $\,$}

\newcommand{\niin}{\mbox{[\ion{N}{2}] $\lambda$6584}}

\newcommand{\myoiiihbn}{\mbox{[\ion{O}{3}]}/{\mbox{H$\beta$}}}
\newcommand{\myniihan}{\mbox{[\ion{N}{2}]}/{\mbox{H$\alpha$}}}
\newcommand{\myniiha}{\mbox{[\ion{N}{2}]}/{\mbox{H$\alpha$} $\,$}}
\newcommand{\siihan}{\mbox{[\ion{S}{2}]}/{\mbox{H$\alpha$}}}
\newcommand{\oihan}{\mbox{[\ion{O}{1}]}/{\mbox{H$\alpha$}}}
\newcommand{\oiiihan}{\mbox{[\ion{O}{3}]}/{\mbox{H$\alpha$}}}

\slugcomment{Accepted for publication in ApJ}
\shortauthors{Comerford et al.}
\shorttitle{}

\begin{document}

\title{A Catalog of 406 AGNs in MaNGA: A Connection between Radio-mode AGN and Star Formation Quenching}

\author{Julia M. Comerford\altaffilmark{1}, James Negus\altaffilmark{1}, Francisco M\"{u}ller-S\'{a}nchez\altaffilmark{2}, Michael Eracleous\altaffilmark{3}, Dominika Wylezalek\altaffilmark{4}, Thaisa Storchi-Bergmann\altaffilmark{5}, Jenny E. Greene\altaffilmark{6}, R. Scott Barrows\altaffilmark{1}, Rebecca Nevin\altaffilmark{1}, Namrata Roy\altaffilmark{7}, and Aaron Stemo\altaffilmark{1}}

\affil{$^1$Department of Astrophysical and Planetary Sciences, University of Colorado, Boulder, CO 80309, USA}

\affil{$^2$Physics Department, University of Memphis, Memphis, TN 38152, USA}

\affil{$^3$Department of Astronomy and Astrophysics and Center for Gravitational Wave Physics, The Pennsylvania State University, 525 Davey Lab, University Park, PA 16803, USA}

\affil{$^4$European Southern Observatory, Karl-Schwarzschildstr 2, D-85748 Garching bei M\"{u}nchen, Germany}

\affil{$^5$Departamento de Astronomia, Universidade Federal do Rio Grande do Sul, IF, CP 15051, 91501-970 Porto Alegre, RS, Brazil}

\affil{$^6$Department of Astrophysical Sciences, Princeton University, Princeton, NJ 08544, USA}

\affil{$^7$Department of Astronomy and Astrophysics, University of California Santa Cruz, 1156 High Street, CA 95064, USA}

\begin{abstract}

Accurate active galactic nucleus (AGN) identifications and spatially resolved host galaxy properties are a powerful combination for studies of the role of AGNs and AGN feedback in the coevolution of galaxies and their central supermassive black holes.  Here, we present robust identifications of 406 AGNs in the first 6261 galaxies observed by the integral field spectroscopy survey Mapping Nearby Galaxies at Apache Point Observatory (MaNGA).  Instead of using optical line flux ratios, which can be difficult to interpret in light of the effects of shocks and metallicity, we identify the AGNs via mid-infrared {\it WISE} colors, {\it Swift}/BAT ultra hard X-ray detections, NVSS and FIRST radio observations, and broad emission lines in SDSS spectra.  We subdivide the AGNs into radio-quiet and radio-mode AGNs, and examine the correlations of the AGN classes with host galaxy star formation rates and stellar populations.  When compared to the radio-quiet AGN host galaxies, we find that the radio-mode AGN host galaxies reside preferentially in elliptical galaxies, lie further beneath the star-forming main sequence (with lower star formation rates at fixed galaxy mass), have older stellar populations, and have more negative stellar age gradients with galactocentric distance (indicating inside-out quenching of star formation).  These results establish a connection between radio-mode AGNs and the suppression of star formation. \\

\end{abstract}  


\section{Introduction}
\label{intro}

One of the most important issues in astrophysics today is understanding how galaxies and their supermassive black holes (SMBHs) coevolve, resulting in surprisingly tight correlations such as the $M_{BH}-\sigma$ relation (e.g., \citealt{GE00.1,GR06.1,MC13.1}; see \citealt{HE14.1} for a review).  Active galactic nuclei (AGNs) have emerged as key players in this coevolution, since SMBH mass growth is traced by nuclear activity, and negative feedback from AGNs can quench star formation in a galaxy and regulate the galaxy's mass growth (e.g., \citealt{SI98.1,DI05.1,HO08.3,FA12.2,ST19.1}).  Negative AGN feedback is invoked not only to explain observed correlations between SMBHs and their host galaxies, but also to produce the observed break in the galaxy luminosity function and the bimodal color distribution of galaxies (e.g., \citealt{ST01.2,SI12.1}).

The role and significance of AGN feedback depends on the class of AGN.  AGNs can be subdivided into two classes: radio-quiet AGNs and radio-loud AGNs.  Radio-quiet AGNs are not radio silent; they have smaller 1.4 GHz radio powers and radio-to-optical flux density ratios ($\simlt 10^{24}$ W Hz$^{-1}$ and $\simlt 10$, respectively; e.g., \citealt{FA74.1,LE96.2}) than radio-loud AGNs.  Radio-quiet AGNs (whether Type 1 or Type 2 AGNs) are quasar-mode AGNs, which are also known as radiative-mode AGNs or high-excitation radio galaxies (HERGs).  Radio-loud AGNs can be further subdivided; they are either quasar-mode AGNs or they are radio-mode AGNs, which are also known as jet-mode AGNs or low-excitation radio galaxies (LERGs; e.g., \citealt{TA08.1,HI09.1,HE14.1}).   Quasar-mode AGNs accrete large amounts of gas from an optically thick, geometrically thin accretion disk that is radiatively efficient (e.g., \citealt{SH73.1}), and most of their energy is emitted radiatively.  In contrast, radio-mode AGNs are thought to accrete material at a lower rate through advection-dominated accretion flows (e.g., \citealt{NA95.3}), which leads to most of the energy being emitted as kinetic energy in powerful radio jets rather than emitted radiatively (e.g., \citealt{ME07.3}).

Even though quasar-mode AGNs contribute two orders of magnitude more energy to the universe than radio-mode AGNs (e.g., \citealt{CA09.1}), radio-mode AGNs are a promising feedback mechanism because most of the radio jet energy can be imparted directly to the host galaxy (e.g., \citealt{WA11.4}).  In contrast, quasar-mode AGNs may be less efficient at imparting their energy to the host galaxy and influencing star formation.

There are two possible ways that radio-mode AGNs can influence their host galaxies: either by quenching star formation (negative feedback) or enhancing star formation (positive feedback).  The negative feedback scenario works by the radio jet's mechanical energy heating the interstellar medium (ISM) in a galaxy, preventing the gas from radiatively cooling and forming stars (e.g., \citealt{PA09.1,NE10.1}) or by driving the molecular gas needed to form stars out of the galaxy (e.g., \citealt{NE06.1,MO10.1}).  In the positive feedback scenario, the radio jets can shock the ISM, creating the high density conditions for the ISM to then collapse and form new stars (e.g., \citealt{KA12.1,ZI13.1}). 

Much remains unclear about the role of AGNs in general, and radio-mode AGNs in particular, in driving feedback in galaxies.  To begin with, our understanding of the influence of AGNs and feedback in galaxy evolution and galaxy - SMBH coevolution is limited by our ability to accurately identify AGNs.  Most spectroscopic surveys identify AGNs using Baldwin-Phillips-Terlevich optical emission line diagnostics (BPT; \citealt{BA81.1,KE01.2}), which classify the emission source as star formation, a low-ionization nuclear emission-line region (LINER), Seyfert, or composite.  However, many studies have illustrated that these BPT diagnostics are a minefield for AGN identifications, since optical lines can be obscured by dust and the emission line ratios can be changed by effects such as shocks, variations in metallicity, young hot stars, and evolved hot stars.  Shock ionization related to AGNs or star formation can move emission line ratios into different regions of the BPT diagram (e.g., \citealt{RI11.1,KE13.1}).  Metallicity can also change emission line ratios; for example, because nitrogen is a secondary element its abundance traces the metallicity, and consequently lower metallicities produce lower \niin / \ha flux ratios (e.g., \citealt{VA98.2,GR06.3}).  Emission from young, hot stars such as Wolf-Rayet stars can elevate line flux ratios in the BPT diagram (e.g., \citealt{BR08.3}), while post-asymptotic giant branch stars are capable of producing hard ionizing radiation that alters BPT line flux ratios (e.g., \citealt{BI94.1,YA12.1,BE16.1}). 

Other measurements can be considered alongside the BPT diagnostics to help mitigate the contamination by non-AGNs.  For example, \ha equivalent widths can help separate ionization by AGNs from ionization by post-AGB stars (e.g., \citealt{CI10.3,KE19.1,SA20.1}), and the distance from the standard diagnostic lines can quantify the significance of a BPT classification \citep{WY18.1}. Reprojections of the BPT diagrams can also resolve ambiguities in the classifications (e.g., \citealt{JI20.1}).  The best path forward to improve BPT-based diagnostics of AGNs is still developing.

More accurate and complete AGN identifications can advance any study of AGNs in galaxies, and are particularly compelling for the SDSS-IV survey Mapping Nearby Galaxies at Apache Point Observatory (MaNGA) since MaNGA enables spatially-resolved studies of galaxies in unprecedented numbers ($\sim 10,000$ nearby galaxies; \citealt{BU15.1,LA15.1}).  Here, we present a catalog of secure identifications of 406 AGNs in MaNGA via four observational approaches, using both ground- and space-based archival data.  Using spatially-resolved properties of the host galaxies, we then assess the relationship between radio-mode AGNs and star formation, which is not yet well understood (e.g., \citealt{MO17.1,WY18.2}).  Hereafter, we refer to negative feedback as `feedback'.

This paper is organized as follows.  In Section~\ref{manga}, we present the parent sample of MaNGA galaxies and their derived properties.  In Section~\ref{agn}, we present the catalog of 406 AGNs, which are selected by mid-infrared colors, ultra hard X-ray detections, radio observations, and broad emission lines.  In Section~\ref{compare}, we compare the overlap of our 406 AGNs with other approaches to selecting AGNs in MaNGA, including BPT-based AGN identifications. In Section~\ref{results}, we present and discuss our results, which establish a connection between radio-mode AGNs and the suppression of star formation in galaxies.  Finally, Section~\ref{conclusions} summarizes our conclusions.

We assume a Hubble constant $H_0 =70$ km s$^{-1}$ Mpc$^{-1}$, $\Omega_m=0.3$, and $\Omega_\Lambda=0.7$ throughout, and all distances are given in physical (not comoving) units.

\section{Galaxy Sample and Properties}
\label{manga}

We base our catalog on the galaxies observed by MaNGA, which is an ongoing SDSS-IV integral field spectroscopy (IFS) survey of $\sim 10,000$ low-redshift galaxies.  MaNGA began taking data in 2014 and will finish in 2020 \citep{BU15.1,DR15.1,LA15.1,YA16.1,BL17.1,WA17.1}.  MaNGA uses IFS with $2^{\prime\prime}$ fibers grouped into hexagonal bundles, and the observations span 3600 - 10,300 \AA \, with a spectral resolving power of $R\sim2000$.  The redshift range is $0.01<z<0.15$ (average redshift $z=0.03$).   The PSF FWHM is $2\farcs5$, which corresponds to physical resolutions of 0.5 kpc to 6.5 kpc for this redshift range.  The hexagonal bundles range in diameter from $12\farcs5$ to $32\farcs5$. 

MaNGA targets galaxies with stellar masses $>10^9$ $M_\odot$. The more massive, elliptical galaxies are typically found at the higher redshifts \citep{BU15.1,WA17.1}, and the galaxy morphologies evolve over the MaNGA redshift range \citep{SA19.1}.  Finally, the MaNGA survey was designed to spectroscopically map galaxies out to at least 1.5 times the effective radius ($R_e$), and the typical MaNGA galaxy is mapped out to a radius of $\sim15$ kpc.  

We use the eighth MaNGA Product Launch (MPL-8) of the data reduction pipeline \citep{LA16.2}, which includes observations of 6261 galaxies observed through mid-2018.  Our analyses rely on galaxy properties measured in the Pipe3D Value Added Catalog \citep{SA16.1,SA18.2}, including galaxy stellar mass, star formation rate (SFR; derived from stellar population modeling), stellar population age (luminosity-weighted age of the stellar population at $R_e$), and stellar age gradient (slope of the gradient of the luminosity-weighted log-age of the stellar population within a galactocentric distance of 0.5 to 2.0 $R_e$).

\section{MaNGA AGN Catalog}
\label{agn}

\subsection{AGN Selection}
\label{selection}

We select AGNs using four different approaches  -- {\it WISE} mid-infrared color cuts, {\it Swift}/BAT hard X-ray sources, NVSS/FIRST 1.4 GHz radio sources, and SDSS broad emission lines -- where each data set has full coverage of the MaNGA footprint.  We outline each approach below.  

\subsubsection{WISE Mid-infrared Colors}
\label{wise}

Mid-infrared emission, which is produced by hot dust in the obscuring structure around the AGN, is a good selector for both obscured and unobscured AGNs.  First, we select AGNs using {\it Wide-field Infrared Survey Explorer} ({\it WISE}) mid-infrared observations of the MaNGA galaxies.  {\it WISE} observed the full sky in four bands at 3.4 $\mu$m, 4.6 $\mu$m, 12 $\mu$m, and 22 $\mu$m ($W1$, $W2$, $W3$, and $W4$, respectively).  We crossmatch the AllWISE Source Catalog \citep{WR10.1,MA14.1} to the MaNGA galaxies with a matching radius of $6\farcs25$ (the smallest radius of a MaNGA integral field unit), and we find 6417 matches.  For these matches, we use the magnitudes measured with profile-fitting photometry and provided by AllWISE.  The $6^{\prime\prime}$ PSF for the $W1$, $W2$, and $W3$ bands corresponds to 2-15 kpc for our redshift range and encompasses the hot dust producing the mid-infrared emission (e.g., \citealt{EL06.3}). 

There are many different {\it WISE} color criteria that have been used to select AGNs (e.g., \citealt{WR10.1,JA11.1,DO12.1,ST12.2,AS13.1,AS18.1}), and we select one with a well-defined, high reliability because our goal is to assemble a catalog of robust AGN identifications.  Reliability is measured as the fraction of {\it WISE} selected AGNs that have AGN contributions of $>50$\% to their spectral energy distributions (see \citealt{AS10.1,AS13.1,AS18.1}). Specifically, we apply the 75\% reliability criteria of $W1-W2>0.486 \, \exp \{ 0.092(W2-13.07)^2 \}$ and $W2>13.07$, or $W1-W2 > 0.486$ and $W2 \leq 13.07$ to select AGNs \citep{AS18.1}, which yields 67 AGNs.   

To determine the bolometric luminosity of each AGN, we estimate the rest-frame 6 $\mu$m luminosity and convert it to the restframe 2-10 keV luminosity \citep{ST15.1}.  We note that there is a range in bolometric corrections and they may depend on Eddington ratios (e.g., \citealt{VA07.1}).   We convert the restframe 2-10 keV luminosity to the bolometric luminosity by multiplying by a factor of 20, which is a typical bolometric correction for AGNs (e.g., \citealt{EL94.1,MA04.4}).  

\subsubsection{Swift/BAT Ultra Hard X-rays}

Hard X-rays are reliable tracers of the hot X-ray corona around AGNs, and hard X-rays are especially valuable for identifying AGNs because these high energy X-rays are less biased by orientation effects or obscuration.  The {\it Swift} observatory's Burst Alert Telescope (BAT) is carrying out a uniform all-sky survey in the ultra hard X-ray (14 - 195 keV), and the recent 105-month BAT catalog has identified $\sim1000$ AGN \citep{OH18.1}.  They crossmatch their X-ray source catalog to Data Release 12 of the Sloan Digital Sky Survey (SDSS DR12; \citealt{AL15.2}), which encompasses all of the MaNGA galaxies that have been observed, and we find that there are 17 BAT-identified AGNs in MaNGA.  

We then convert the published 14 - 195 keV luminosities to bolometric luminosities using the bolometric correction $L_{bol}/L_{14-195 \, \mathrm{keV}}=8.47$ \citep{RI17.2,IC19.1}.

\subsubsection{NVSS/FIRST 1.4 GHz Radio Sources}
\label{radio}

Radio observations are useful for detecting the radio jet emission associated with AGNs.  Observations from the NRAO Very Large Array Sky Survey (NVSS; \citealt{CO98.3}) and the Faint Images of the Radio Sky at Twenty centimeters (FIRST; \citealt{BE95.1}) have been used to identify AGNs in SDSS DR7 galaxies \citep{BE12.1}.  NVSS is a 1.4 GHz continuum survey that fully covers the sky north of a declination of -$40$ deg, while FIRST is a 1.4 GHz survey of 10,000 square degrees of the North and South Galactic Caps.   The \cite{BE12.1} catalog has a flux density limit of 5 mJy, which extends down to a 1.4 GHz radio luminosity of $\sim 10^{23}$ W Hz$^{-1}$ at $z=0.1$.  They separate emission from AGNs from emission from star formation by using the correlation between the 4000\AA \, break strength and the radio luminosity per stellar mass \citep{BE05.3}, BPT diagnostics, and the correlation between the H$\alpha$ luminosity and the radio luminosity; see \cite{BE12.1} for more details.  They estimate that $\simlt1\%$ of objects are misclassified \citep{BE05.3}.  Since all MaNGA galaxies are also SDSS DR7 galaxies, we crossmatch the \cite{BE12.1} DR7 AGN catalog with MaNGA and find 325 radio AGNs in MaNGA.

We convert the 1.4 GHz integrated fluxes of the sources to 2-10 keV luminosities via the scaling relation in \cite{PA15.1}, and then apply a bolometric correction as described in Section~\ref{wise}.

Further, \cite{BE12.1} have identified which AGNs are quasar-mode (or high-excitation radio galaxies; HERGs) and which AGNs are radio-mode (or low-excitation radio galaxies; LERGs; we note that LERGs are also closely related to the weak-line radio galaxies of \citealt{TA98.1}).  They base this classification on the emission lines detected in the the SDSS DR7 spectra of the galaxies.  Their classification follows four main approaches: (1) using the excitation index parameter of \cite{BU10.1}, which is based on the emission line ratios \myoiiihbn, \myniihan, \siihan, and \oihan; (2) using the \cite{KE06.1} diagnostic diagram; (3) using the \oiii equivalent width; and (4) using the \cite{CI10.3} diagnostic, which is based on the emission line ratios \myniiha and \oiiihan.  The full details of the classification are presented in \cite{BE12.1}.  We find only three radio-loud quasar-mode AGNs in our sample, which is unsurprising given that  MaNGA focuses on lower luminosity AGNs in general. 

We compare the radio-mode AGNs to the radio-quiet AGNs in Section~\ref{results}.

\subsubsection{Broad Emission Lines}

Broad Balmer emission lines are excellent tracers of Type 1 AGNs, as the broad lines are produced in the high density gas very close to the SMBH (e.g., \citealt{OS91.1}).  \cite{OH15.1} recently analyzed the spectra of SDSS DR7 galaxies at $z<0.2$ for evidence of broad H$\alpha$ emission lines, which they used to build a catalog of Type 1 AGNs in SDSS.    We crossmatch their catalog with MaNGA, since all MaNGA galaxies are included in SDSS DR7, and we find 55 broad-line AGNs in MaNGA.

Then, we convert the \oiiiw luminosities published in \cite{OH15.1} to bolometric luminosities via the scaling relation of \cite{PE17.1}.

\subsection{The MaNGA AGN Catalog}
\label{catalog}

In total we identify 406 unique AGNs in MaNGA MPL-8, and we present the AGNs, how they were identified, their bolometric luminosities, and whether they are quasar-mode or radio-mode (for the radio AGNs) in a catalog here, with the data fields as defined in Table~\ref{tbl-1}.   

We identify the AGNs using {\it WISE} colors, BAT X-ray detections, radio observations, and broad emission lines, and each approach has its own strengths and limitations; 46 of the AGNs are identified by more than one approach, as shown in Table~\ref{tbl-2}.  {\it WISE} mid-infrared colors are excellent for uncovering even obscured AGNs, but there can also be significant, contaminating mid-infrared emission from the host galaxy.  As a result, mid-infrared colors can be biased towards AGNs that are accreting at high rates (e.g., \citealt{HI09.1,ME13.2}).  BAT ultra hard X-rays can identify AGNs over a greater range of accretion rates, but are less sensitive to extremely obscured, Compton thick AGNs (column densities $N_H \geq 10^{24}$ cm$^{-2}$;  e.g., \citealt{AK16.1,MA18.1}).  While radio observations can penetrate through this dust to detect obscured AGNs, they have difficulty identifying low-intensity, radio-quiet AGNs (e.g., \citealt{WH00.2}).   Finally, broad emission lines are robust indicators of Type 1 AGNs, but they can be diluted by the stellar continuum and they miss all obscured (Type 2) AGNs.

Every selector of AGNs has its caveats.  Since the limitations of one AGN selector can be overcome by the strengths of another AGN selector, a multiwavelength approach that uses many different selectors -- such as the approach followed here -- can build the most complete sample of AGNs.

We find that each AGN subsample has a similar median host galaxy mass:  $\log (M_* / M_\odot)$=[11.2, 11.0, 11.0, 10.9] for the radio, broad-line, {\it WISE}, and BAT subsamples, respectively.

In Section~\ref{compare} we compare our MaNGA AGN catalog to other AGN catalogs assembled for MaNGA, while in Section~\ref{results} we focus on comparisons between the host galaxies of the 81 radio-quiet AGNs (those identified by {\it WISE}, BAT, and/or broad emission lines, but undetected in radio) and the host galaxies of the 143 radio-mode AGNs, to better understand the relationship between these types of AGNs and star formation.

\begin{deluxetable*}{lll}
\tablewidth{0pt}
\tablecolumns{3}
\tablecaption{Data Fields in the MaNGA AGN Catalog} 
\tablehead{
\colhead{No.} &
\colhead{Field} & 
\colhead{Description} 
}
\startdata 
1 & MANGA\_ID & Galaxy identifier assigned by MaNGA \\
2 & RA & Right ascension of MaNGA galaxy [J2000, decimal degrees] \\
3 & DEC & Declination of MaNGA galaxy [J2000, decimal degrees] \\
4 & Z & Spectroscopic redshift of galaxy \\
5 & WISE\_AGN & Whether AGN was selected in {\it WISE} [Boolean] \\
6 & LOG\_LBOL\_WISE & Log bolometric luminosity of {\it WISE} AGN [erg s$^{-1}$] \\
7 & LOG\_LBOL\_WISE\_ERR & Log bolometric luminosity error of {\it WISE} AGN [erg s$^{-1}$] \\
8 & BAT\_AGN & Whether AGN was selected in BAT [Boolean] \\
9 & LOG\_LBOL\_BAT & Log bolometric luminosity of BAT AGN [erg s$^{-1}$] \\
10 & LOG\_LBOL\_BAT\_ERR & Log bolometric luminosity error of BAT AGN [erg s$^{-1}$] \\
11 & RADIO\_AGN & Whether AGN was selected in NVSS/FIRST [Boolean] \\
12 & RADIO\_CLASS & Quasar-mode (HERG) or radio-mode (LERG) \\
13 & LOG\_LBOL\_RADIO & Log bolometric luminosity of NVSS/FIRST AGN [erg s$^{-1}$] \\
14 & LOG\_LBOL\_RADIO\_ERR & Log bolometric luminosity error of NVSS/FIRST AGN [erg s$^{-1}$] \\
15 & BROAD\_AGN & Whether AGN was selected by broad lines [Boolean]  \\
16 & LOG\_LBOL\_BROAD & Log bolometric luminosity of broad-line AGN [erg s$^{-1}$] \\
17 & LOG\_LBOL\_BROAD\_ERR & Log bolometric luminosity error of broad-line AGN [erg s$^{-1}$] 
\enddata
\tablecomments{The MaNGA AGN Catalog is available in its entirety in fits format from the original publisher.}
\label{tbl-1}
\end{deluxetable*}

\begin{deluxetable}{lr} 
\tablewidth{0pt}
\tablecolumns{2}
\tablecaption{Overlap between AGN Classifications} 
\tablehead{
\colhead{AGN Identifier(s)} &
\colhead{Number of AGNs} 
}
\startdata 
{\it WISE} only & 25 \\
BAT only & 1 \\
Radio only & 309 \\
Broad-lines only & 25 \\ 
{\it WISE} \& BAT & 3 \\ 
{\it WISE} \& radio & 11 \\ 
{\it WISE} \& broad & 18 \\ 
{\it WISE}, BAT, \& radio & 2 \\ 
{\it WISE}, BAT, \& broad-lines & 7 \\ 
{\it WISE}, BAT, radio, \& broad-lines & 1 \\ 
BAT \& broad-lines & 2 \\ 
BAT, radio, \& broad-lines & 1 \\ 
Radio \& broad-lines & 1
\enddata
\label{tbl-2}
\end{deluxetable}

\section{Comparison to Other AGN Catalogs in MaNGA} 
\label{compare}

Here we compare our 406 AGNs to other studies that have selected AGNs or AGN candidates in MaNGA.  Our catalog is unique in that it is dominated by radio AGNs, and we are particularly interested in the degree of overlap between our sample and the AGN samples that were selected by combinations of BPT line flux ratios, H$\alpha$ equivalent width, and surface brightness (Section~\ref{rembold}; Section~\ref{wylezalek}; Section~\ref{sanchez}).  The $2^{\prime\prime}$ angular size of the MaNGA fibers corresponds to 1.2 kpc at the average redshift of $z=0.03$, which means that even the spectrum of a galaxy's nuclear spaxel can encompass a range of energy sources besides an AGN.  The BPT and H$\alpha$ selected AGNs that do not appear in our catalog are candidates for misclassifications due to shocks, metallicity, young hot stars, and evolved hot stars; this will be the topic of a subsequent paper.

\subsection{MaNGA Ancillary AGN Catalog}
\label{ancillary}

While in its planning stages, MaNGA added several luminous AGN host galaxies to its target list to increase the range of AGN luminosities that MaNGA samples (ancillary AGN program; PI: J. Greene)\footnote{\tt https://www.sdss.org/dr15/manga/manga-target-selection/ ancillary-targets/luminous-agn/}.  This ancillary AGN program increased the maximum AGN luminosity in MaNGA from $L_{bol} \sim 10^{43}$ erg s$^{-1}$ to $L_{bol} \sim 10^{45}$ erg s$^{-1}$.  The goal of the program was to select Type 2 AGNs, so that both the spatially-extended narrow-line region and the host galaxy properties could be studied.  These additional AGNs were selected using the {\it Swift}/BAT catalog of hard X-ray sources, \oiiiw fluxes from the AGN Line Profile and Kinematics Archive \citep{MU13.3}, and the {\it WISE} color cuts of $0.7 < W1-W2 < 2.0$ and $2.0 < W2-W3 < 4.5$ \citep{WR10.1} as well as a minimum bolometric luminosity of $10^{43}$ erg s$^{-1}$ to help exclude star-forming galaxies.  The program also ensured that each ancillary AGN had at least two inactive MaNGA control galaxies of similar redshift, stellar mass, and size.

In SDSS DR15, which includes the 4656 galaxies in MaNGA MPL-7, 24 of the ancillary AGNs were observed.  Of these, two were selected for the ancillary AGN catalog by BAT alone, 12 were selected by \oiiiw alone, nine were selected by {\it WISE} alone, and one was selected by both BAT and \oiiiwn.  Eight of the ancillary AGNs are included, by one or more of our diagnostics (Section~\ref{selection}), in our MaNGA AGN catalog.  Of the remaining 16 ancillary AGNs that are {\it not} in our MaNGA AGN catalog, nine are \oiiiw selected AGNs.  Since we did not use optical emission lines in our AGN selection, it makes sense that not all of the \oiiiw selected AGNs would also be in our sample.  The final seven ancillary AGNs that are in the ancillary catalog but not in ours are {\it WISE} selected AGNs.  The reason for the different classifications is the different {\it WISE} color cut criteria; we have selected our criteria to be purposefully conservative (Section~\ref{wise}), which omits these ancillary AGNs.

\subsection{Rembold et al. MaNGA AGN Catalog}
\label{rembold}

\cite{RE17.1} searched for AGNs in MaNGA using both BPT and WHAN (where \ha equivalent widths $>3$ \AA \, identify ionization by AGNs, and not ionization by post-AGB stars; \citealt{CI10.3}) diagnostics simultaneously.  Using the SDSS-III integrated nuclear spectra of the 2778 galaxies observed in MaNGA MPL-5, they identify 62 `true' AGNs whose line flux ratios and \ha equivalent widths lie in the Seyfert or LINER regions of both the BPT and WHAN diagrams

We identify 13 of the 62 AGNs in our catalog (21\%): six are detected by radio only, three are detected by {\it WISE} and radio, two are detected by broad lines only, one is detected by {\it WISE} only, and one is detected by {\it WISE} and BAT.  

\subsection{Wylezalek et al. MaNGA AGN Catalog}
\label{wylezalek}

\cite{WY18.1} made the first large systematic effort to use spatially-resolved BPT diagrams to identify AGN candidates with MaNGA data.  For each MaNGA galaxy, they created a BPT map where each spaxel's ionizing radiation is classified as star formation, LINER, Seyfert, or composite.  They identified AGN candidates as those with a certain fraction of the spaxels having line flux ratios consistent with Seyferts, as well as using cuts on H$\alpha$ surface brightness and equivalent width. From these criteria, they selected 308 AGN candidates out of the first 2727 galaxies that were observed by MaNGA in MPL-5.  

\cite{WY18.1} note that their objects are AGN {\it candidates}, because there are many known contaminants that can mimic AGN-like signatures in photoionized gas.  For example, the line flux ratios can be affected by shocks, metallicity, obscuration, young hot stars, evolved hot stars, and off-nuclear AGN (e.g., \citealt{RI11.1,KE13.1,BE16.1}).

Of their 308 AGN candidates, 41 are in our AGN catalog (13\%): 18 are detected by radio only, six are detected by {\it WISE} only, six are detected by broad lines only, five are detected by {\it WISE} and radio, four are detected by {\it WISE} and broad lines, one is detected by {\it WISE} and BAT, and one is detected by {\it WISE}, BAT, and broad lines.  

\begin{figure}[!t]
\centering
\includegraphics[width=9cm]{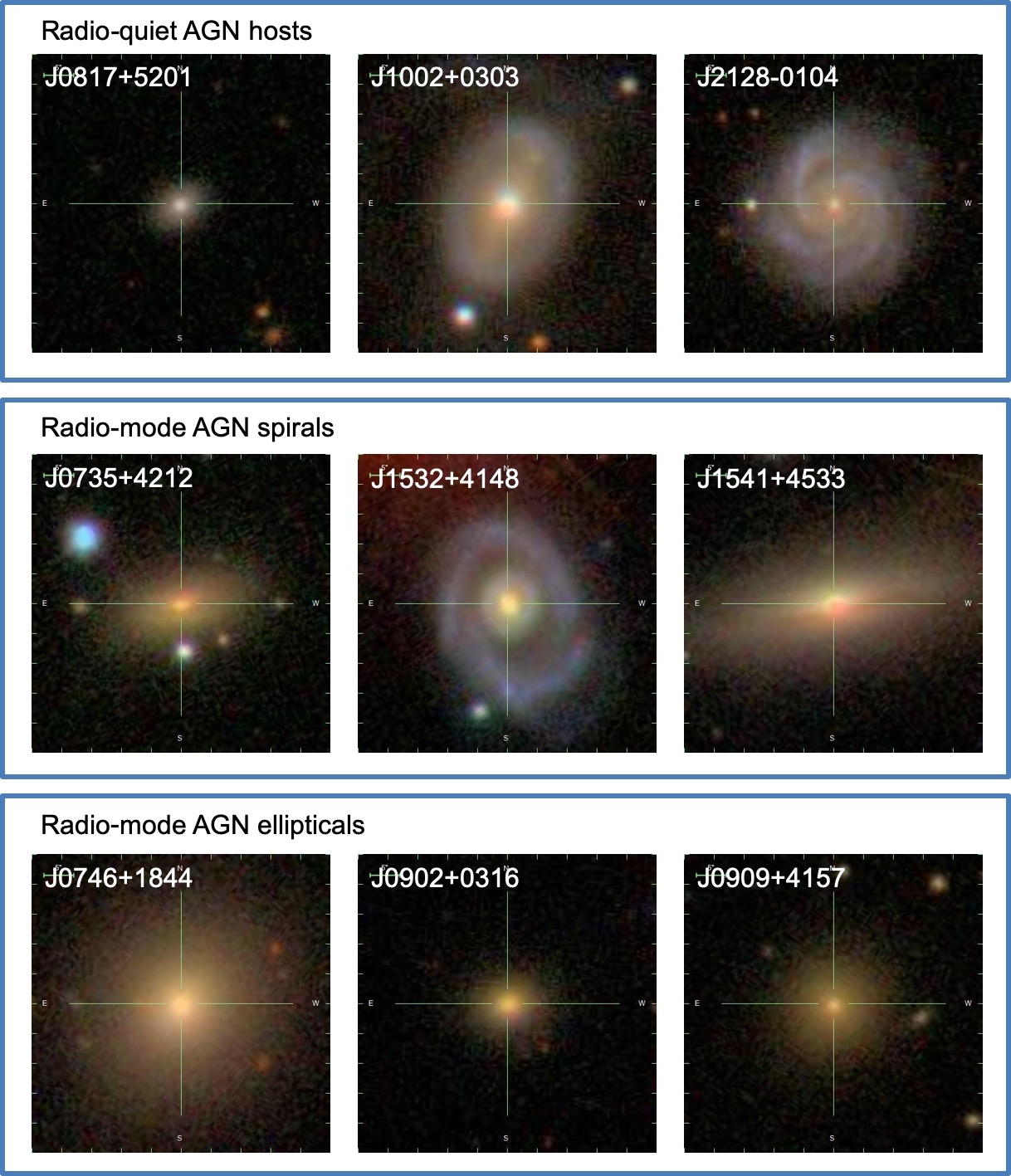}
\caption{Example galaxy morphologies for the radio-quiet AGN host galaxies (top row), radio-mode AGN spiral galaxies (middle row), and radio-mode AGN elliptical galaxies (bottom row).  Each panel is a $50^{\prime\prime} \times 50^{\prime\prime}$ cutout of the SDSS $ugriz$ image. }
\label{fig:morph}
\end{figure}

\subsection{S\'{a}nchez et al. MaNGA AGN Catalog}
\label{sanchez}

\cite{SA18.2} selected AGN candidates based on the integrated spectra of the central $3^{\prime\prime}$ by $3^{\prime\prime}$ of MaNGA galaxies.  To classify an object as an AGN candidate, they required that this central integrated spectrum have emission line ratios that lie above the theoretical maximum for starbursts in the BPT diagram \citep{KE01.2, KE06.1} and an H$\alpha$ equivalent width that is $> 1.5 \, \AA \,$ \citep{CI10.3}.  Out of 2755 galaxies in MaNGA MPL-5, they identify 98 AGN candidates.

We find that 23 of their AGNs are also in our AGN catalog (23\%): eight are detected by radio only, three are detected by {\it WISE} only, six are detected by broad lines only, one is detected by {\it WISE} and radio, three are detected by {\it WISE} and broad lines, one is detected by {\it WISE} and BAT, and one is detected by radio, broad lines, and BAT.

The catalogs of \cite{RE17.1}, \cite{WY18.1}, and \cite{SA18.2} are similar in that they each employ variations of BPT and H$\alpha$ measurements to select AGN candidates in MaNGA.  We find that $\sim10-20\%$ of their AGN candidates are recovered in our catalog.  We will explore the physical reasons for these overlaps, and the implications for using BPT diagnostics and H$\alpha$ measurements to select AGNs, in a subsequent paper (Negus et al., in prep.).

\section{Results and Discussion}
\label{results}

\begin{deluxetable}{lll}
\tablewidth{0pt}
\tablecolumns{3}
\tablecaption{Classifications of Odd Host Galaxies of Radio-mode AGNs} 
\tablehead{
\colhead{SDSS name} &
\colhead{Galaxy Zoo} & 
\colhead{Visual} \\ 
\colhead{} &
\colhead{Classification$^a$} &
\colhead{Classification$^b$}
}
\startdata 
SDSS J074351.25+282128.0 & none & merger \\
SDSS J074949.43+345302.1 & merger & merger \\
SDSS J075909.96+294651.7 & merger & merger \\
SDSS J080028.00+413938.2 & none & merger \\
SDSS J081141.12+360656.8 & none & merger \\
SDSS J081343.61+525738.2 & something else & merger or lens \\
SDSS J082133.17+550907.0 & merger & merger \\
SDSS J084453.99+274308.3 & none & merger \\
SDSS J090234.90+204417.9 & merger & merger \\
SDSS J110941.19+214425.3 & something else & merger \\
SDSS J114316.27+551639.6 & something else & unknown \\
SDSS J121039.44+363652.1 & something else & merger \\
SDSS J145558.28+323732.5 & something else & merger \\
SDSS J151554.87+344346.3 & merger & merger \\
SDSS J153227.65+414842.3 & ring & spiral with ring \\
SDSS J153929.67+443854.4 & n/a & merger \\
SDSS J161835.15+294902.3 & merger & merger \\
SDSS J162441.33+251941.6 & merger & merger \\
SDSS J162650.63+255328.2 & something else & merger \\
SDSS J162823.30+435727.3 & none &  merger
\enddata
\tablenotetext{a}{Where the Galaxy Zoo 2 classification had a weighted vote fraction $>0.5$.  ``None" indicates that no vote fraction exceeded 0.5, while ``n/a" indicates that the galaxy was not included in Galaxy Zoo 2.}
\tablenotetext{b}{Based on bye-eye classification of SDSS composite image.}
\label{tbl-3}
\end{deluxetable}

\begin{figure*}[!t]
\centering
\includegraphics[width=17cm]{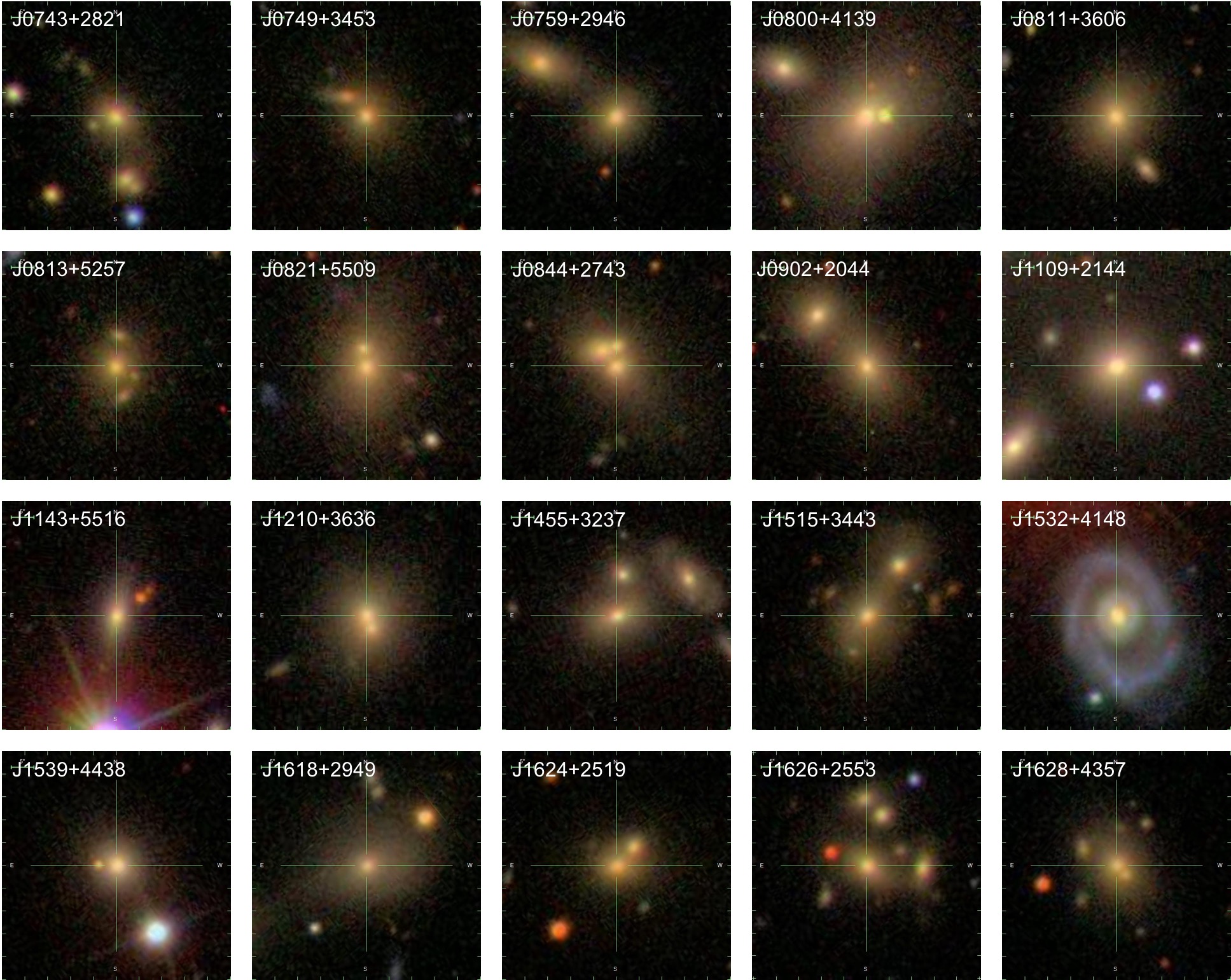}
\caption{$50^{\prime\prime} \times 50^{\prime\prime}$ cutouts of SDSS $ugriz$ images of the radio-mode AGNs whose host galaxies are classified as odd by Galaxy Zoo 2.}
\label{fig:odd}
\end{figure*}

\subsection{Radio-mode AGNs Are Preferentially Found in Elliptical Galaxies and Galaxy Mergers}
\label{morphology}

To understand how radio-mode and radio-quiet AGNs might affect their host galaxies differently, we first compare their host galaxy morphologies.  For morphologies, we use the MaNGA value-added catalog of published morphologies from Galaxy Zoo 2 \citep{WI13.1} and currently unpublished morphologies from Galaxy Zoo 4.  Following \cite{WI13.1}, we identify elliptical galaxies as those classified as ``smooth" and spiral galaxies (a category that includes S0 galaxies) as those classified as ``features or disk".  Galaxy Zoo volunteers are also asked whether there is anything odd about each galaxy, and then whether the odd feature is due to a ring, a lens or arc, a disturbed galaxy, an irregular galaxy, something else, a merger, or a dust lane.

We find that the radio-quiet AGNs are hosted mainly in spiral galaxies. Of the 81 radio-quiet AGNs, 56 are hosted in spirals (69\%), 8 are hosted in ellipticals (10\%), and 0 are hosted in mergers (0\%).  In contrast, the radio-mode AGNs are hosted mainly in elliptical galaxies. Of the 143 radio-mode AGNs, 25 are hosted in spirals (17\%), 93 are hosted in ellipticals (65\%), and $\geq7$ are hosted in mergers ($\geq$5\%; see below).  Figure~\ref{fig:morph} shows examples of the host galaxies of the radio-quiet AGNs and the radio-mode AGNs.  In particular, many of the radio-mode AGNs are found in S0 galaxies that are red; we explore this more in Section~\ref{sfms}, where we find that the radio-mode AGN host galaxies indeed have low SFRs.

Our findings confirm previous studies that have established that radio-mode AGNs preferentially reside in massive, red, elliptical galaxies (e.g., \citealt{BA92.3,VE01.2,KA08.1,SM09.3,JA12.1,HA18.1}).  These elliptical galaxies must have central nuclear gas reservoirs to accrete onto the central SMBHs and ignite the AGNs.

The radio-mode AGNs are also unique in that 20 (14\%) of them are hosted in galaxies that are marked as odd, whereas 0 of the radio-quiet AGN host galaxies are classified as odd.  The 20 galaxies with odd morphologies are shown in Figure~\ref{fig:odd}, and Table~\ref{tbl-3} shows the reason for why each galaxy is classified as odd (where we only list the classifications with weighted vote fractions $>0.5$).  The most common classification is ``merger", with 7 (35\%) of the odd galaxies receiving this classification.  As another check, we visually classified the 20 odd galaxies and found that 17 (85\%) of them are mergers interacting with one or more nearby companions (Table~\ref{tbl-3}).  As Figure~\ref{fig:odd} shows, most of these mergers appear to be between elliptical galaxies.

Several studies have also found that, in mass-matched samples, radio-loud AGNs are more likely to be found in galaxy mergers than radio-quiet AGNs (e.g., \citealt{RA12.1,CH15.2}).   These specific radio-mode AGNs are tracers of gas inflow triggering nuclear activity in mergers.

Our finding, that 7 to 17 of the 143 radio-mode AGNs in MaNGA (5-12\%) are hosted by galaxy mergers, is lower than the 27\% of $z<0.7$ weak-line radio galaxies that have disturbed, merger-like morphologies \citep{RA11.1}.  This difference may be explained by the depth of the \cite{RA11.1} Gemini optical broad-band observations (features detected at a median depth of $\mu_V=23.6$ mag arcsec$^{-2}$), whereas the Galaxy Zoo classifications were made from SDSS images with a median depth $\mu_r=22.7$ mag arcsec$^{-2}$.  Many more of the MaNGA galaxies may have faint post-merger signatures that are only discernible with deeper follow-up imaging.  

Radio-mode AGNs might be found preferentially in elliptical galaxies because they could be a last phase in the evolution of AGNs; the host galaxies may have slowed their star formation due to processes such as morphological quenching that are unrelated to AGNs (e.g., \citealt{MA09.1}).  Alternatively, radio-mode AGNs might be found preferentially in elliptical galaxies because radio-mode AGNs drive feedback in their host galaxies, quenching star formation.  Another possibility that could also apply to the mergers is that interactions or other AGN triggering events in elliptical galaxies, where there is little gas available to fuel the AGNs, lead to low accretion rates to the SMBHs, producing radio-mode AGNs.  To explore the effects of the radio-mode AGNs on their host galaxies, including whether they might drive feedback, we turn to the SFRs of the host galaxies. 

\subsection{AGNs, and Especially Radio-mode AGNs, Lie Below the Star-forming Main Sequence}
\label{sfms} 

On average, galaxies form stars at a rate commensurate with the mass of the galaxy; this is seen in the observed correlation between the SFR and stellar mass $\mathrm{M}_*$ of galaxies, which is sometimes called the ``star-forming main sequence" (SFMS; \citealt{BR04.1,EL07.1,NO07.1,SP14.1}).  The existence of a SFMS suggests that star formation may be limited by both the internal gas supply and by quenching mechanism(s) that exhaust the available gas supply for new star formation.

\begin{figure}
\centering
\includegraphics[height=8cm]{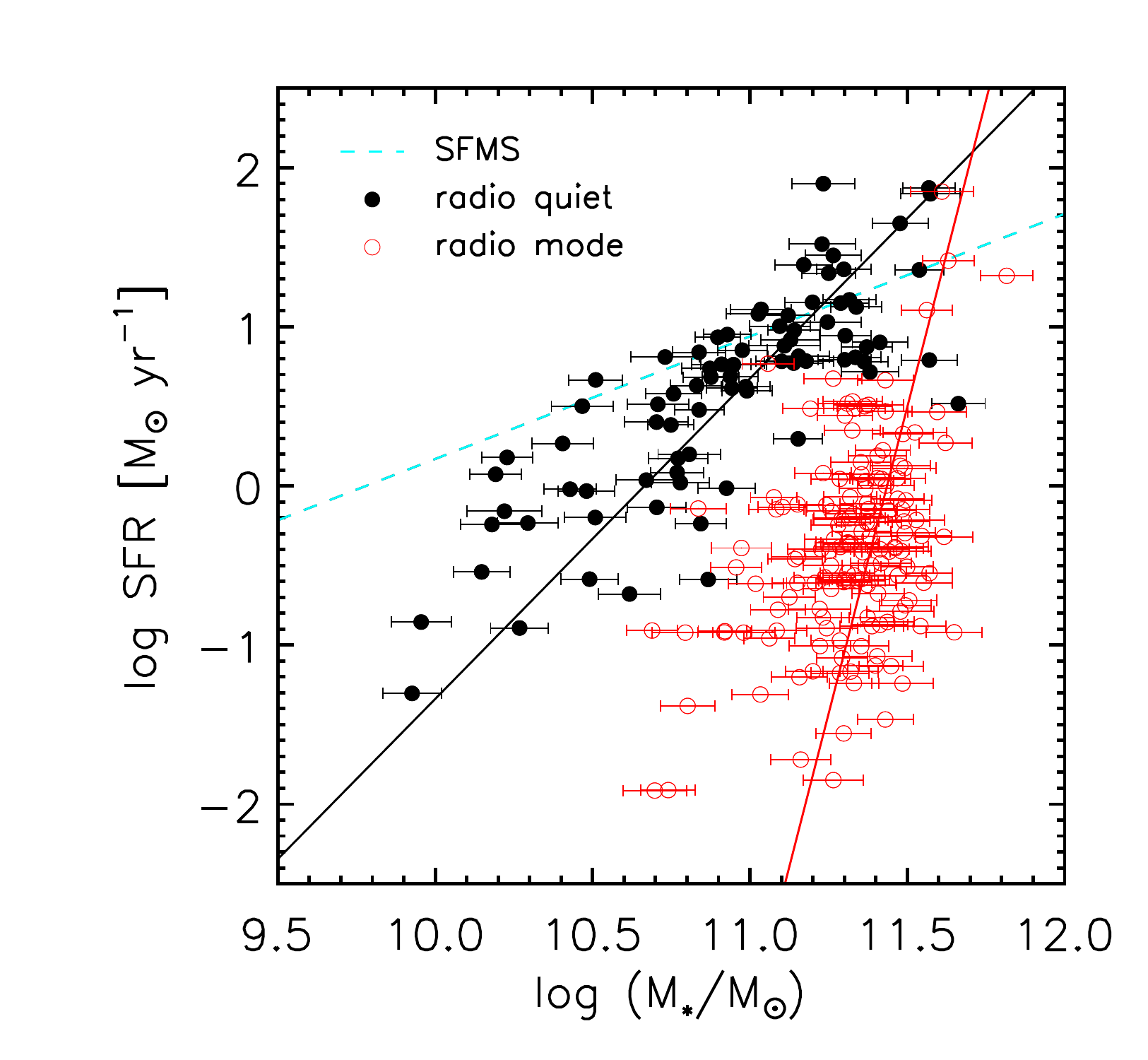} 
\caption{Relation between SFR and stellar mass for radio-quiet AGN host galaxies (filled black circles; best-fit relation shown as black solid line) and radio-mode AGN host galaxies (open red circles; best-fit relation shown as red solid line).  The error bars on SFR are too small to see, with a median error of 0.005 on $\log \mathrm{(SFR / M_\odot \, yr^{-1})}$.  For reference, the dashed cyan line illustrates the star-forming main sequence for local SDSS galaxies with blue colors \citep{EL07.1}.}
\label{fig:sfms}
\end{figure} 

\begin{figure}
\centering
\includegraphics[width=8.5cm]{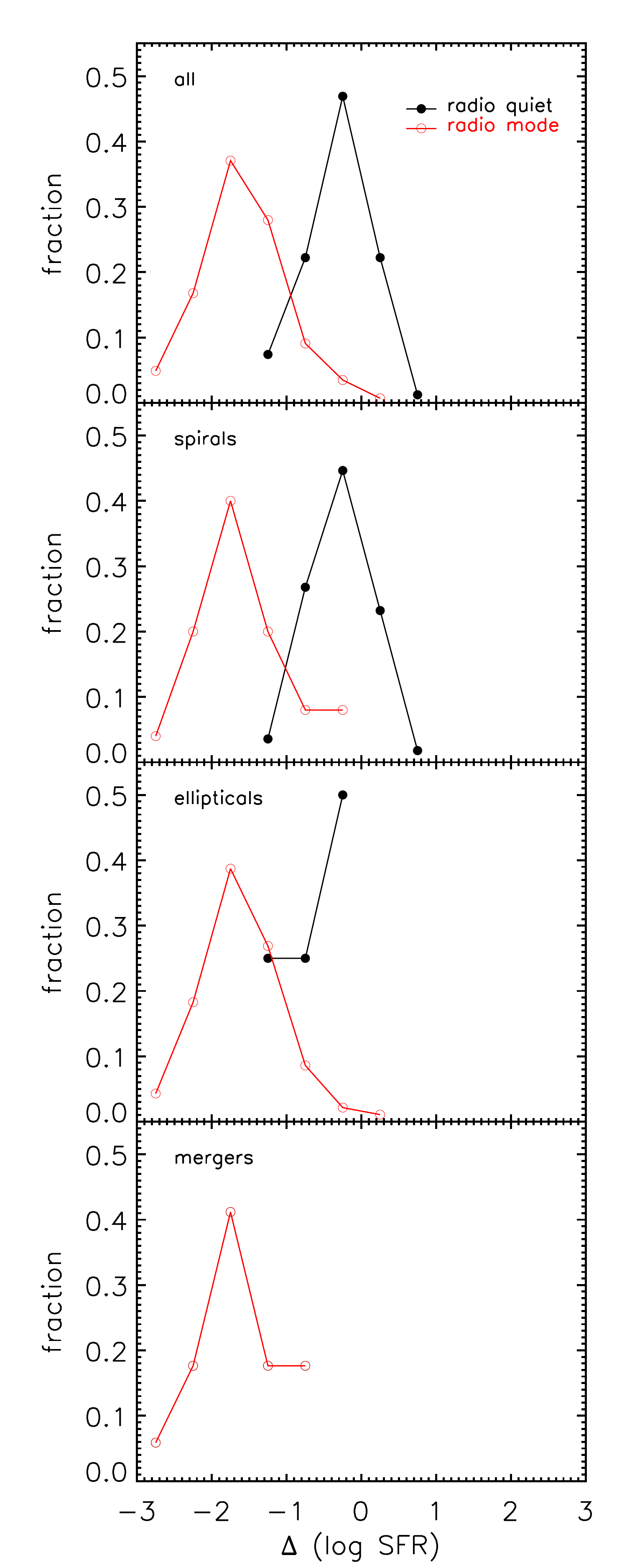} 
\caption{Fraction of AGN host galaxies versus distance from the star-forming main sequence, where $\Delta (\log \mathrm{SFR})= \log \mathrm{SFR}_{AGN} - \log \mathrm{SFR}_{SFMS}$, shown for different host galaxy morphologies.  Regardless of host galaxy morphology, the radio-mode AGN host galaxies (open red circles) lie significantly further below the SFMS than the radio-quiet AGN host galaxies (filled black circles).  There are no radio-quiet AGN host galaxies that are classified as mergers.  We have binned the data with bin sizes of $\Delta (\log \mathrm{SFR})=0.5$.}
\label{fig:delta}
\end{figure}

\begin{figure}
\centering
\includegraphics[width=8.5cm]{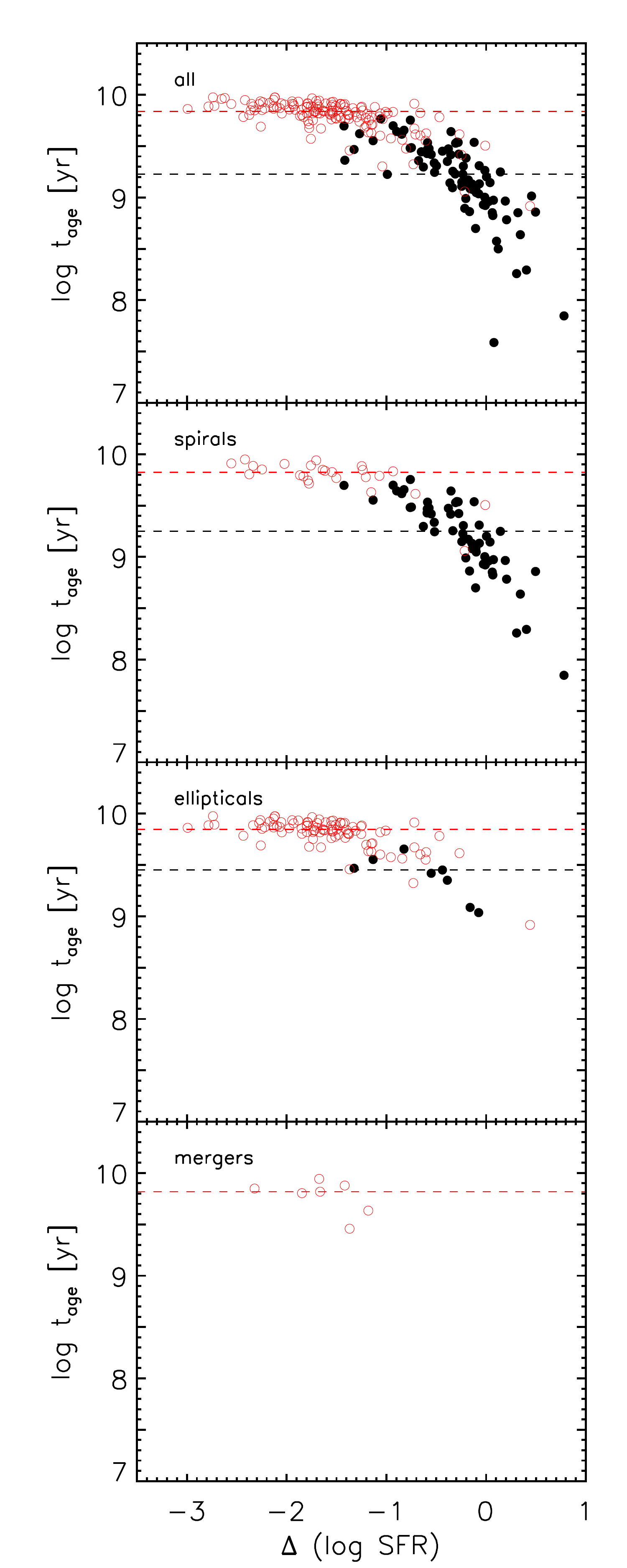} 
\caption{Stellar population ages as a function distance from the star-forming main sequence, where $\Delta (\log \mathrm{SFR})= \log \mathrm{SFR}_{AGN} - \log \mathrm{SFR}_{SFMS}$, shown for different host galaxy morphologies.  The radio-quiet AGN host galaxies  are shown as filled black circles (with median stellar population ages shown as the black dashed lines), while the radio-mode AGN host galaxies are shown as open red circles (with median stellar populations age shown as the red dashed lines).  For all host galaxy morphologies, the radio-mode AGN host galaxies lie further beneath the SFMS and have older stellar populations (approaching the age of the universe), suggesting that recent star formation has been shut down in these galaxies.}
\label{fig:ages}
\end{figure}

However, the presence of AGNs complicates the picture, as some studies find that AGN host galaxies lie predominantly along the SFMS (e.g., \citealt{RO13.1,ST17.1}); some find that AGN host galaxies lie above the SFMS, with enhanced SFRs compared to those of inactive galaxies of similar mass (e.g., \citealt{YO14.1,PI16.1}); while still others find that AGN host galaxies lie under the SFMS, with suppressed SFRs compared to those of inactive galaxies of similar mass (e.g., \citealt{MU15.2,SH15.2}). If AGN host galaxies lie above or below the SFMS, this suggests that the AGN itself may be linked to the SFR -- for example, by the AGN enhancing or suppressing star formation. In general AGN host galaxies are found in the green valley, between star-forming and elliptical galaxies (e.g., \citealt{SC10.1,SH17.1,SA18.2}), but the SFR of a particular host galaxy may depend on the type of AGN that it hosts (e.g., \citealt{EL16.1}). 

To investigate this further, we plot the SFR and galaxy stellar mass of the radio-quiet and the radio-mode AGN host galaxies in MaNGA (Figure~\ref{fig:sfms}).  For comparison, we use the SFMS that was derived for local ($0.015 \leq z \leq 0.1$) SDSS galaxies with blue optical colors \citep{EL07.1}. The first difference we notice in AGN host galaxies is that the radio-mode AGNs are preferentially found in higher mass galaxies, which is consistent with our finding in Section~\ref{morphology} that radio-mode AGNs are preferentially found in elliptical galaxies (see also, e.g., \citealt{MA64.1, JA12.1}).

For each population of AGN host galaxies (radio-quiet, and radio-mode), we fit the SFR and stellar mass with a power law
\begin{equation}
\log \mathrm{(SFR / M_\odot \, yr^{-1})}= \alpha + \beta \log (M_* / M_\odot) \; .
\end{equation}
We find that the best-fit relation for the radio-quiet AGN host galaxies ($\alpha=-21.5 \pm 0.7$, $\beta=2.01 \pm 0.06$) is steeper than the SFMS, with the majority of AGN host galaxies falling below the SFMS (systematically lower SFR for a given stellar mass).  To quantify this, we calculate each AGN host galaxy's distance from the SFMS as $\Delta (\log \mathrm{SFR})= \log \mathrm{SFR}_{AGN} - \log \mathrm{SFR}_{SFMS}$ (Figure~\ref{fig:delta}).  For the radio-quiet AGN host galaxies, the median $\Delta (\log \mathrm{SFR})$ is $-0.2$.  This is similar to the median $\Delta (\log \mathrm{SFR}) \sim -0.5$ found for {\it Swift}/BAT AGNs at $z<0.05$ as well as BPT-selected AGNs at $z<0.08$ \citep{SH15.2}.  Other studies of SDSS galaxies at $0.04 < z < 0.1$ find that BPT-selected AGNs lie $\sim 0$ to 2 dex below the SFMS \citep{LE16.1,MC19.1}.

The radio-mode AGN host galaxies are even more extreme; they exhibit a steeper SFR-$M_*$ relation ($\alpha=-88.1 \pm 8.1$, $\beta=7.7 \pm 0.7$) and lie even further beneath the SFMS (median $\Delta (\log \mathrm{SFR})$ of $-1.6$) than the radio-quiet AGN host galaxies, as shown in Figure~\ref{fig:sfms} and Figure~\ref{fig:delta}.  Since radio-mode AGNs are often found in elliptical galaxies (Section~\ref{morphology}), it follows that they should lie below the SFMS on average. Our results are also in general agreement with the radio-excess-selected AGNs that show lower SFRs for a given galaxy mass \citep{KA14.1}, although a more direct comparison is not possible because that study's AGN selection (via fits to the spectral energy distribution) and redshift range ($z<2$) are not comparable to this work.  

We also divide the radio-mode AGNs and radio-quiet AGNs by host galaxy morphology, and find that the same trend holds: regardless of host galaxy morphology, the radio-mode AGN host galaxies fall further beneath the SFMS than the radio-quiet AGN host galaxies (Figure~\ref{fig:delta}).  These results are particularly interesting for elliptical galaxies, since they are gas-poor in general.  However, the elliptical galaxies that host AGNs must have nuclear gas available to accrete onto their SMBHs.  This gas could also form stars, but since we are finding lower SFRs in these galaxies, this implies that the AGNs (and particularly the radio-mode AGNs) could be quenching star formation.

We find that AGNs, and particularly radio-mode AGNs, reside in host galaxies with suppressed star formation.  We explore this further via the stellar properties of the MaNGA host galaxies.

\begin{deluxetable}{lllll}
\tablewidth{0pt}
\tablecolumns{5}
\tablecaption{Median Stellar Age Gradients} 
\tablehead{
\colhead{ } &
\multicolumn{4}{c}{$\log (M_*/M_\odot)$} \\
\colhead{ } &
\colhead{10.9} &
\colhead{11.1} &
\colhead{11.3} &
\colhead{11.5}
}
\startdata 
radio-quiet AGNs & & & & \\
\hspace{0.1in} spiral & -0.08 & 0.04 & -0.11 & -0.15 \\
\hspace{0.1in} elliptical & 0.28 & n/a & 0.20 & n/a \\
\hline
radio-mode AGNs & & & & \\
\hspace{0.1in} spiral & n/a & -0.16 & -0.06 & -0.13 \\
\hspace{0.1in} elliptical & -0.15 & -0.11 & -0.08 & -0.06 \\
\hspace{0.1in} merger & -0.19 & -0.14 & -0.10 & -0.24
\enddata
\tablecomments{Host galaxy stellar mass bins have size 0.2 dex and are centered on the stellar mass given by each column header. Bins that have no galaxies in them are marked with n/a.}
\label{tbl-4}
\end{deluxetable}

\subsection{Radio-mode AGN Host Galaxies Have Older Stellar Populations and Negative Stellar Age Gradients}

After finding that the radio-mode AGN host galaxies lie significantly under the star-forming main sequence, which shows that star formation is suppressed in these galaxies, we now search for further evidence of quenching in the host galaxies' stellar population ages and age gradients.

First, we consider the stellar population age $t_{age}$ (defined as the luminosity-weighted age of the stellar population at the galaxy's effective radius; Section~\ref{manga}) for each host galaxy.  As Figure~\ref{fig:ages} shows, the radio-mode AGN host galaxies not only lie further beneath the SFMS but also have stellar population ages that are older than those of the radio-quiet AGN host galaxies.  This is true independent of host galaxy morphology, though it is a somewhat stronger effect for spiral galaxies.  We find that the median stellar ages of the radio-quiet AGN host galaxies are $\log (t_{age}/\mathrm{yr}) = [9.2, 9.4]$ for spiral and elliptical galaxies, respectively, while the median stellar ages of the radio-mode AGN host galaxies are $\log (t_{age}/\mathrm{yr}) = [9.8, 9.8]$ for spiral and elliptical galaxies, respectively.

Similarly, \cite{CH13.1} studied a sample of SDSS galaxies at $0.1 < z < 0.3$ with stellar masses $\log (M_*/M_\odot) > 11.4$ and found that the fraction of galaxies that have had star formation within the last Gyr is lower in galaxies with radio jets than in radio-quiet galaxies.  Our finding that radio-mode AGN host galaxies have stellar populations that are a median of 4-5 Gyr older than the stellar populations of radio-quiet AGN host galaxies indicates that star formation has been quenched in radio-mode AGN host galaxies in the past.  We explore this idea further with stellar age gradients.

Since MaNGA obtains spatially-resolved spectra across the galaxies that it observes, we can measure how the stellar population ages change as a function of galactocentric radius across a given galaxy.  Here, we define the stellar age gradient $\alpha$ as the slope of the gradient of the luminosity-weighted log-age of the stellar population within a galactocentric distance of 0.5 to 2.0 $R_e$ (Section~\ref{manga}).  When $\alpha$ is positive, it means that the stellar populations are getting older as we move away from the galactic center.  A negative $\alpha$ indicates that the stellar populations are getting younger as we move away from the galactic center.  We compare the stellar age gradients of the radio-quiet AGN host galaxies and the radio-mode AGN host galaxies to look for signs of quenching in either type of galaxy.

\begin{figure} 
\centering
\includegraphics[height=8cm]{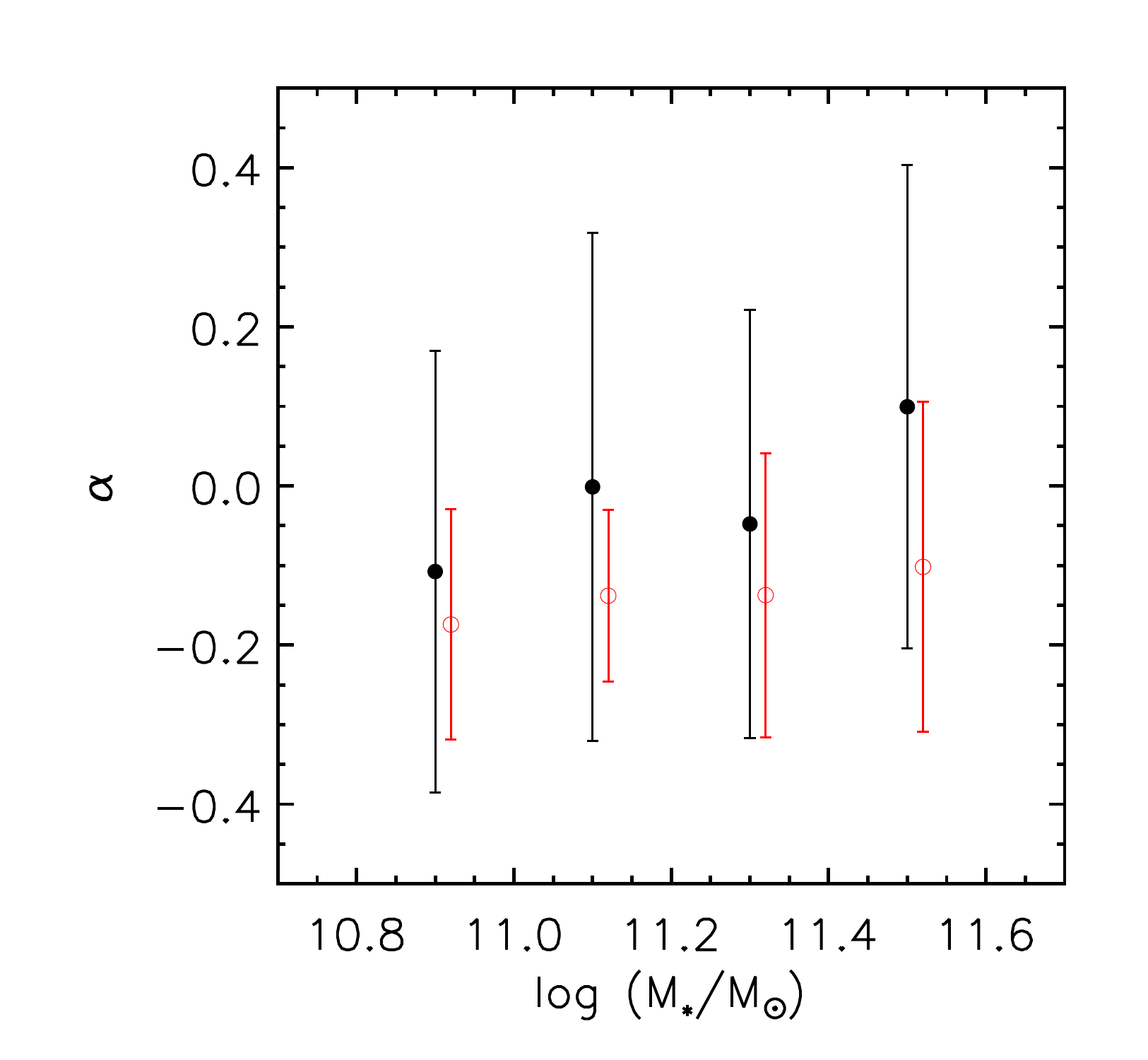}
\caption{Stellar age gradients $\alpha$ for the radio-quiet AGN host galaxies (filled black circles) and radio-mode AGN host galaxies (open red circles, offset in the x-axis for clarity). The mean stellar age gradients are plotted against stellar mass, in stellar mass bins of size 0.2 dex.  The error bars represent the standard deviation in the $\alpha$ values.  While the radio-quiet AGN host galaxies show flat stellar age gradients across each galaxy ($\alpha \sim 0$), the radio-mode AGN host galaxies show negative ($\alpha \sim -0.15$) stellar age gradients; this suggests inside-out quenching of star formation in the radio-mode AGN host galaxies.}
\label{fig:gradients}
\end{figure}

Because stellar age gradients correlate with host galaxy stellar mass (e.g., \citealt{GO14.1,ZH17.1}), we compare the stellar age gradients in stellar mass bins of 0.2 dex.  As Figure~\ref{fig:gradients} shows, the radio-quiet AGN host galaxies have $\sim$flat ($\alpha \sim 0$) stellar age gradients while the radio-mode AGN host galaxies have negative ($\alpha \sim -0.15$) stellar age gradients.  For the radio-mode AGN host galaxies, this means that the central stellar populations are older than the stellar populations further towards the galaxy outskirts.  

It has been shown that the stellar age gradient measurement depends on the number of fibers in the integral field unit, with lower numbers of fibers leading to overestimates and higher numbers of fibers leading to more accurate measurements \citep{IB19.1}.  To determine the magnitude of this effect on our results, we reanalyze the stellar age gradients for only the galaxies with the largest fiber bundle size (127 fibers, corresponding to a diameter of $32\farcs5$).   We find that the stellar age gradients decrease by $\sim 0.05$; the radio-quiet AGN host galaxies have $\alpha \sim -0.05$ while the radio-mode AGN host galaxies have $\alpha \sim -0.2$.  The overall trend, of radio-mode AGN host galaxies exhibiting more negative stellar age gradients than radio-quiet AGN host galaxies, remains the same.

We also compare the stellar age gradients in different host galaxy morphologies, binned by stellar mass.  The median stellar age gradients for the radio-quiet AGNs and radio-mode AGNs, subdivided by host galaxy morphology and host galaxy stellar mass, are shown in Table~\ref{tbl-4}.  The radio-quiet AGN host galaxies exhibit negative stellar age gradients for most of the spiral galaxy mass bins, and positive stellar age gradients for the elliptical galaxies.  In contrast, the radio-mode AGN host galaxies exhibit negative stellar age gradients for all morphology types.

This result, that radio-mode AGN host galaxies of all masses and all morphologies have negative stellar age gradients -- even when the radio-quiet AGN host galaxies have positive stellar age gradients (as in the case of elliptical galaxies) -- implies inside-out suppression of new star formation in the host galaxies of radio-mode AGNs. This could be explained by AGN feedback evacuating the gas and preventing central star formation. This same mechanism could also starve the central SMBH of gas, leading the SMBH to accrete at a lower rate via the advection-dominated accretion flows (e.g., \citealt{IS14.1}) that create the radio-mode AGNs that are currently observed in these galaxies. The reason for the galaxy mergers to have more negative stellar age gradients than the spiral and elliptical galaxies may be that the mergers themselves induce additional inside-out suppression of star formation (e.g., \citealt{BA17.2}).  An alternative explanation is that radio-mode AGNs, which are associated with low accretion rates onto SMBHs, occur preferably in elliptical galaxies with older stellar populations in the nuclear region, due to the lack or exhaustion of gas to form new stars.

\section{Conclusions}
\label{conclusions}

We have assembled a catalog of 406 unique AGNs in MaNGA MPL-8, which consists of observations of 6261 galaxies.  Instead of using optical emission line diagnostics, which are particularly prone to misclassifications, we identify the AGNs by their {\it WISE} colors, {\it Swift}/BAT ultra hard X-ray sources, NVSS/FIRST radio observations, or broad emission lines.  We compare the radio-mode AGNs (some of which were also identified as AGNs by {\it WISE}, BAT, and/or broad lines) to the population of radio-quiet AGNs that were identified as AGNs by {\it WISE}, BAT, and/or broad lines.  Then, we search for correlations between AGNs and the star formation in their host galaxies, by using MaNGA observations to examine the AGN host galaxy locations on the star-forming main sequence and the stellar population ages and age gradients of the AGN host galaxies. 

Our main results are summarized below.

\vspace{.1in}

1.  Radio-quiet AGNs are hosted primarily in spiral galaxies (88\% of radio-quiet AGNs are found in spirals), while radio-mode AGNs are predominantly found in elliptical galaxies (74\% of radio-mode AGNs are found in ellipticals).  While no radio-quiet AGNs were found to be in mergers, 5-12\% of radio-mode AGNs reside in mergers. \\

2. AGN host galaxies lie below the star-forming main sequence of SFR versus galaxy stellar mass; for fixed stellar mass, the AGN host galaxies have systematically lower SFRs.  The radio-quiet AGN host galaxies are found at a median $\Delta (\log \mathrm{SFR}) = -0.2$, while the radio-mode AGN host galaxies lie even further below the star-forming main sequence at a median $\Delta (\log \mathrm{SFR}) = -1.6$ (Figures~\ref{fig:sfms},~\ref{fig:delta}). \\

3.  Radio-mode AGN host galaxies have stellar populations that are a median of 4-5 Gyr older than the stellar populations of radio-quiet AGN host galaxies (Figure~\ref{fig:ages}).  This is independently true for each host galaxy morphology type. \\

4.  Overall, radio-quiet AGN host galaxies have flat stellar population age gradients, while radio-mode AGN host galaxies have negative ($\alpha \sim -0.15$) stellar age gradients (Figure~\ref{fig:gradients}).  In elliptical galaxies, radio-quiet AGNs have positive stellar age gradients (median $\alpha=0.20$) and radio-mode AGNs have negative stellar age gradients (median $\alpha=-0.09$).  Our finding that radio-mode AGNs are associated with negative stellar age gradients means that radio-mode AGN host galaxies have older stellar populations in their centers and younger stellar populations at larger distances from the galaxy centers.  This is evidence of inside-out quenching of star formation in these radio-mode AGN host galaxies. \\

5.  When the previous four results are considered together (radio-mode AGN host galaxies are typically elliptical, lie far below the star-forming main sequence, have older stellar populations, and have negative stellar age gradients), they show a connection between radio-mode AGNs and star formation quenching in the past, particularly in galaxy centers.  The cause of the connection is not clear.  Possibilities include: (1) radio-mode AGN feedback is quenching star formation; (2) radio-mode AGNs are a final phase in the evolution of AGNs; (3) interactions or other gas transfer events in elliptical galaxies, where there is little gas available to fuel the AGNs, lead to low accretion rates onto the SMBHs, consequently producing radio-mode AGNs. \\

Further evidence of AGN feedback in these MaNGA galaxies could be obtained by searching for outflows in these galaxies and then modeling the outflows to determine their energetics, to constrain the impact on the star-forming material.  Observations of the molecular gas content of the host galaxies could also determine if gas is being removed from the central regions of the galaxies, which would suppress central star formation.

\acknowledgements J.M.C. is supported by NSF AST-1714503. We thank the anonymous referee for comments that have improved the clarity of this paper.

Funding for the Sloan Digital Sky Survey IV has been provided by the Alfred P. Sloan Foundation, the U.S. Department of Energy Office of Science, and the Participating Institutions. SDSS-IV acknowledges support and resources from the Center for High-Performance Computing at the University of Utah. The SDSS web site is www.sdss.org.

SDSS-IV is managed by the Astrophysical Research Consortium for the  Participating Institutions of the SDSS Collaboration including the Brazilian Participation Group, the Carnegie Institution for Science, Carnegie Mellon University, the Chilean Participation Group, the French Participation Group, Harvard-Smithsonian Center for Astrophysics, Instituto de Astrof\'isica de Canarias, The Johns Hopkins University, Kavli Institute for the Physics and Mathematics of the Universe (IPMU) / University of Tokyo, the Korean Participation Group, Lawrence Berkeley National Laboratory, Leibniz Institut f\"ur Astrophysik Potsdam (AIP),  Max-Planck-Institut f\"ur Astronomie (MPIA Heidelberg), Max-Planck-Institut f\"ur Astrophysik (MPA Garching), Max-Planck-Institut f\"ur Extraterrestrische Physik (MPE), National Astronomical Observatories of China, New Mexico State University, New York University, University of Notre Dame, Observat\'ario Nacional / MCTI, The Ohio State University, Pennsylvania State University, Shanghai Astronomical Observatory, United Kingdom Participation Group, Universidad Nacional Aut\'onoma de M\'exico, University of Arizona, University of Colorado Boulder, University of Oxford, University of Portsmouth, University of Utah, University of Virginia, University of Washington, University of Wisconsin, Vanderbilt University, and Yale University.  This project makes use of the MaNGA-Pipe3D dataproducts. We thank the IA-UNAM MaNGA team for creating this catalogue, and the Conacyt Project CB-285080 for supporting them.

\bibliographystyle{apj}

\begin{thebibliography}{121}
\expandafter\ifx\csname natexlab\endcsname\relax\def\natexlab#1{#1}\fi

\bibitem[{{Akylas} {et~al.}(2016){Akylas}, {Georgantopoulos}, {Ranalli},
  {Gkiokas}, {Corral}, \& {Lanzuisi}}]{AK16.1}
{Akylas}, A., {Georgantopoulos}, I., {Ranalli}, P., {Gkiokas}, E., {Corral},
  A., \& {Lanzuisi}, G. 2016, \aap, 594, A73

\bibitem[{{Alam} {et~al.}(2015){Alam}, {Albareti}, {Allende Prieto}, {Anders},
  {Anderson}, {Anderton}, {Andrews}, {Armengaud}, {Aubourg}, {Bailey}, \&
  et~al.}]{AL15.2}
{Alam}, S., {Albareti}, F.~D., {Allende Prieto}, C., {Anders}, F., {Anderson},
  S.~F., {Anderton}, T., {Andrews}, B.~H., {Armengaud}, E., {Aubourg}, {\'E}.,
  {Bailey}, S., \& et~al. 2015, \apjs, 219, 12

\bibitem[{{Assef} {et~al.}(2010){Assef}, {Kochanek}, {Brodwin}, {Cool},
  {Forman}, {Gonzalez}, {Hickox}, {Jones}, {Le Floc'h}, {Moustakas}, {Murray},
  \& {Stern}}]{AS10.1}
{Assef}, R.~J., {Kochanek}, C.~S., {Brodwin}, M., {Cool}, R., {Forman}, W.,
  {Gonzalez}, A.~H., {Hickox}, R.~C., {Jones}, C., {Le Floc'h}, E.,
  {Moustakas}, J., {Murray}, S.~S., \& {Stern}, D. 2010, \apj, 713, 970

\bibitem[{{Assef} {et~al.}(2013){Assef}, {Stern}, {Kochanek}, {Blain},
  {Brodwin}, {Brown}, {Donoso}, {Eisenhardt}, {Jannuzi}, {Jarrett}, {Stanford},
  {Tsai}, {Wu}, \& {Yan}}]{AS13.1}
{Assef}, R.~J., {Stern}, D., {Kochanek}, C.~S., {Blain}, A.~W., {Brodwin}, M.,
  {Brown}, M.~J.~I., {Donoso}, E., {Eisenhardt}, P.~R.~M., {Jannuzi}, B.~T.,
  {Jarrett}, T.~H., {Stanford}, S.~A., {Tsai}, C.-W., {Wu}, J., \& {Yan}, L.
  2013, \apj, 772, 26

\bibitem[{{Assef} {et~al.}(2018){Assef}, {Stern}, {Noirot}, {Jun}, {Cutri}, \&
  {Eisenhardt}}]{AS18.1}
{Assef}, R.~J., {Stern}, D., {Noirot}, G., {Jun}, H.~D., {Cutri}, R.~M., \&
  {Eisenhardt}, P.~R.~M. 2018, \apjs, 234, 23

\bibitem[{{Baldwin} {et~al.}(1981){Baldwin}, {Phillips}, \&
  {Terlevich}}]{BA81.1}
{Baldwin}, J.~A., {Phillips}, M.~M., \& {Terlevich}, R. 1981, \pasp, 93, 5

\bibitem[{{Barrows} {et~al.}(2017){Barrows}, {Comerford}, {Zakamska}, \&
  {Cooper}}]{BA17.2}
{Barrows}, R.~S., {Comerford}, J.~M., {Zakamska}, N.~L., \& {Cooper}, M.~C.
  2017, \apj, 850, 27

\bibitem[{{Baum} {et~al.}(1992){Baum}, {Heckman}, \& {van Breugel}}]{BA92.3}
{Baum}, S.~A., {Heckman}, T.~M., \& {van Breugel}, W. 1992, \apj, 389, 208

\bibitem[{{Becker} {et~al.}(1995){Becker}, {White}, \& {Helfand}}]{BE95.1}
{Becker}, R.~H., {White}, R.~L., \& {Helfand}, D.~J. 1995, \apj, 450, 559

\bibitem[{{Belfiore} {et~al.}(2016){Belfiore}, {Maiolino}, {Maraston},
  {Emsellem}, {Bershady}, {Masters}, {Yan}, {Bizyaev}, {Boquien}, {Brownstein},
  {Bundy}, {Drory}, {Heckman}, {Law}, {Roman-Lopes}, {Pan}, {Stanghellini},
  {Thomas}, {Weijmans}, \& {Westfall}}]{BE16.1}
{Belfiore}, F., {Maiolino}, R., {Maraston}, C., {Emsellem}, E., {Bershady},
  M.~A., {Masters}, K.~L., {Yan}, R., {Bizyaev}, D., {Boquien}, M.,
  {Brownstein}, J.~R., {Bundy}, K., {Drory}, N., {Heckman}, T.~M., {Law},
  D.~R., {Roman-Lopes}, A., {Pan}, K., {Stanghellini}, L., {Thomas}, D.,
  {Weijmans}, A.-M., \& {Westfall}, K.~B. 2016, \mnras, 461, 3111

\bibitem[{{Best} \& {Heckman}(2012)}]{BE12.1}
{Best}, P.~N., \& {Heckman}, T.~M. 2012, \mnras, 421, 1569

\bibitem[{{Best} {et~al.}(2005){Best}, {Kauffmann}, {Heckman}, \&
  {Ivezi{\'c}}}]{BE05.3}
{Best}, P.~N., {Kauffmann}, G., {Heckman}, T.~M., \& {Ivezi{\'c}}, {\v Z}.
  2005, \mnras, 362, 9

\bibitem[{{Binette} {et~al.}(1994){Binette}, {Magris}, {Stasi{\'n}ska}, \&
  {Bruzual}}]{BI94.1}
{Binette}, L., {Magris}, C.~G., {Stasi{\'n}ska}, G., \& {Bruzual}, A.~G. 1994,
  \aap, 292, 13

\bibitem[{{Blanton} {et~al.}(2017){Blanton}, {Bershady}, {Abolfathi},
  {Albareti}, {Allende Prieto}, {Almeida}, {Alonso-Garc{\'\i}a}, {Anders},
  {Anderson}, {Andrews}, \& et~al.}]{BL17.1}
{Blanton}, M.~R., {Bershady}, M.~A., {Abolfathi}, B., {Albareti}, F.~D.,
  {Allende Prieto}, C., {Almeida}, A., {Alonso-Garc{\'\i}a}, J., {Anders}, F.,
  {Anderson}, S.~F., {Andrews}, B., \& et~al. 2017, \aj, 154, 28

\bibitem[{{Brinchmann} {et~al.}(2004){Brinchmann}, {Charlot}, {White},
  {Tremonti}, {Kauffmann}, {Heckman}, \& {Brinkmann}}]{BR04.1}
{Brinchmann}, J., {Charlot}, S., {White}, S.~D.~M., {Tremonti}, C.,
  {Kauffmann}, G., {Heckman}, T., \& {Brinkmann}, J. 2004, \mnras, 351, 1151

\bibitem[{{Brinchmann} {et~al.}(2008){Brinchmann}, {Kunth}, \&
  {Durret}}]{BR08.3}
{Brinchmann}, J., {Kunth}, D., \& {Durret}, F. 2008, \aap, 485, 657

\bibitem[{{Bundy} {et~al.}(2015){Bundy}, {Bershady}, {Law}, {Yan}, {Drory},
  {MacDonald}, {Wake}, {Cherinka}, {S{\'a}nchez-Gallego}, {Weijmans}, {Thomas},
  {Tremonti}, {Masters}, {Coccato}, {Diamond-Stanic}, {Arag{\'o}n-Salamanca},
  {Avila-Reese}, {Badenes}, {Falc{\'o}n-Barroso}, {Belfiore}, {Bizyaev},
  {Blanc}, {Bland-Hawthorn}, {Blanton}, {Brownstein}, {Byler}, {Cappellari},
  {Conroy}, {Dutton}, {Emsellem}, {Etherington}, {Frinchaboy}, {Fu}, {Gunn},
  {Harding}, {Johnston}, {Kauffmann}, {Kinemuchi}, {Klaene}, {Knapen},
  {Leauthaud}, {Li}, {Lin}, {Maiolino}, {Malanushenko}, {Malanushenko}, {Mao},
  {Maraston}, {McDermid}, {Merrifield}, {Nichol}, {Oravetz}, {Pan}, {Parejko},
  {Sanchez}, {Schlegel}, {Simmons}, {Steele}, {Steinmetz}, {Thanjavur},
  {Thompson}, {Tinker}, {van den Bosch}, {Westfall}, {Wilkinson}, {Wright},
  {Xiao}, \& {Zhang}}]{BU15.1}
{Bundy}, K., {Bershady}, M.~A., {Law}, D.~R., {Yan}, R., {Drory}, N.,
  {MacDonald}, N., {Wake}, D.~A., {Cherinka}, B., {S{\'a}nchez-Gallego}, J.~R.,
  {Weijmans}, A.-M., {Thomas}, D., {Tremonti}, C., {Masters}, K., {Coccato},
  L., {Diamond-Stanic}, A.~M., {Arag{\'o}n-Salamanca}, A., {Avila-Reese}, V.,
  {Badenes}, C., {Falc{\'o}n-Barroso}, J., {Belfiore}, F., {Bizyaev}, D.,
  {Blanc}, G.~A., {Bland-Hawthorn}, J., {Blanton}, M.~R., {Brownstein}, J.~R.,
  {Byler}, N., {Cappellari}, M., {Conroy}, C., {Dutton}, A.~A., {Emsellem}, E.,
  {Etherington}, J., {Frinchaboy}, P.~M., {Fu}, H., {Gunn}, J.~E., {Harding},
  P., {Johnston}, E.~J., {Kauffmann}, G., {Kinemuchi}, K., {Klaene}, M.~A.,
  {Knapen}, J.~H., {Leauthaud}, A., {Li}, C., {Lin}, L., {Maiolino}, R.,
  {Malanushenko}, V., {Malanushenko}, E., {Mao}, S., {Maraston}, C.,
  {McDermid}, R.~M., {Merrifield}, M.~R., {Nichol}, R.~C., {Oravetz}, D.,
  {Pan}, K., {Parejko}, J.~K., {Sanchez}, S.~F., {Schlegel}, D., {Simmons}, A.,
  {Steele}, O., {Steinmetz}, M., {Thanjavur}, K., {Thompson}, B.~A., {Tinker},
  J.~L., {van den Bosch}, R.~C.~E., {Westfall}, K.~B., {Wilkinson}, D.,
  {Wright}, S., {Xiao}, T., \& {Zhang}, K. 2015, \apj, 798, 7

\bibitem[{{Buttiglione} {et~al.}(2010){Buttiglione}, {Capetti}, {Celotti},
  {Axon}, {Chiaberge}, {Macchetto}, \& {Sparks}}]{BU10.1}
{Buttiglione}, S., {Capetti}, A., {Celotti}, A., {Axon}, D.~J., {Chiaberge},
  M., {Macchetto}, F.~D., \& {Sparks}, W.~B. 2010, \aap, 509, A6

\bibitem[{{Cattaneo} \& {Best}(2009)}]{CA09.1}
{Cattaneo}, A., \& {Best}, P.~N. 2009, \mnras, 395, 518

\bibitem[{{Chen} {et~al.}(2013){Chen}, {Kauffmann}, {Heckman}, {Tremonti},
  {White}, {Guo}, {Wake}, {Schneider}, \& {Schawinski}}]{CH13.1}
{Chen}, Y.-M., {Kauffmann}, G., {Heckman}, T.~M., {Tremonti}, C.~A., {White},
  S., {Guo}, H., {Wake}, D., {Schneider}, D.~P., \& {Schawinski}, K. 2013,
  \mnras, 429, 2643

\bibitem[{{Chiaberge} {et~al.}(2015){Chiaberge}, {Gilli}, {Lotz}, \&
  {Norman}}]{CH15.2}
{Chiaberge}, M., {Gilli}, R., {Lotz}, J.~M., \& {Norman}, C. 2015, \apj, 806,
  147

\bibitem[{{Cid Fernandes} {et~al.}(2010){Cid Fernandes}, {Stasi{\'n}ska},
  {Schlickmann}, {Mateus}, {Vale Asari}, {Schoenell}, \& {Sodr{\'e}}}]{CI10.3}
{Cid Fernandes}, R., {Stasi{\'n}ska}, G., {Schlickmann}, M.~S., {Mateus}, A.,
  {Vale Asari}, N., {Schoenell}, W., \& {Sodr{\'e}}, L. 2010, \mnras, 403, 1036

\bibitem[{{Condon} {et~al.}(1998){Condon}, {Cotton}, {Greisen}, {Yin},
  {Perley}, {Taylor}, \& {Broderick}}]{CO98.3}
{Condon}, J.~J., {Cotton}, W.~D., {Greisen}, E.~W., {Yin}, Q.~F., {Perley},
  R.~A., {Taylor}, G.~B., \& {Broderick}, J.~J. 1998, \aj, 115, 1693

\bibitem[{{Di Matteo} {et~al.}(2005){Di Matteo}, {Springel}, \&
  {Hernquist}}]{DI05.1}
{Di Matteo}, T., {Springel}, V., \& {Hernquist}, L. 2005, \nat, 433, 604

\bibitem[{{Donley} {et~al.}(2012){Donley}, {Koekemoer}, {Brusa}, {Capak},
  {Cardamone}, {Civano}, {Ilbert}, {Impey}, {Kartaltepe}, {Miyaji}, {Salvato},
  {Sanders}, {Trump}, \& {Zamorani}}]{DO12.1}
{Donley}, J.~L., {Koekemoer}, A.~M., {Brusa}, M., {Capak}, P., {Cardamone},
  C.~N., {Civano}, F., {Ilbert}, O., {Impey}, C.~D., {Kartaltepe}, J.~S.,
  {Miyaji}, T., {Salvato}, M., {Sanders}, D.~B., {Trump}, J.~R., \& {Zamorani},
  G. 2012, \apj, 748, 142

\bibitem[{{Drory} {et~al.}(2015){Drory}, {MacDonald}, {Bershady}, {Bundy},
  {Gunn}, {Law}, {Smith}, {Stoll}, {Tremonti}, {Wake}, {Yan}, {Weijmans},
  {Byler}, {Cherinka}, {Cope}, {Eigenbrot}, {Harding}, {Holder}, {Huehnerhoff},
  {Jaehnig}, {Jansen}, {Klaene}, {Paat}, {Percival}, \& {Sayres}}]{DR15.1}
{Drory}, N., {MacDonald}, N., {Bershady}, M.~A., {Bundy}, K., {Gunn}, J.,
  {Law}, D.~R., {Smith}, M., {Stoll}, R., {Tremonti}, C.~A., {Wake}, D.~A.,
  {Yan}, R., {Weijmans}, A.~M., {Byler}, N., {Cherinka}, B., {Cope}, F.,
  {Eigenbrot}, A., {Harding}, P., {Holder}, D., {Huehnerhoff}, J., {Jaehnig},
  K., {Jansen}, T.~C., {Klaene}, M., {Paat}, A.~M., {Percival}, J., \&
  {Sayres}, C. 2015, \aj, 149, 77

\bibitem[{{Elbaz} {et~al.}(2007){Elbaz}, {Daddi}, {Le Borgne}, {Dickinson},
  {Alexander}, {Chary}, {Starck}, {Brandt}, {Kitzbichler}, {MacDonald},
  {Nonino}, {Popesso}, {Stern}, \& {Vanzella}}]{EL07.1}
{Elbaz}, D., {Daddi}, E., {Le Borgne}, D., {Dickinson}, M., {Alexander}, D.~M.,
  {Chary}, R.-R., {Starck}, J.-L., {Brandt}, W.~N., {Kitzbichler}, M.,
  {MacDonald}, E., {Nonino}, M., {Popesso}, P., {Stern}, D., \& {Vanzella}, E.
  2007, \aap, 468, 33

\bibitem[{{Elitzur}(2006)}]{EL06.3}
{Elitzur}, M. 2006, NewAR, 50, 728

\bibitem[{{Ellison} {et~al.}(2016){Ellison}, {Teimoorinia}, {Rosario}, \&
  {Mendel}}]{EL16.1}
{Ellison}, S.~L., {Teimoorinia}, H., {Rosario}, D.~J., \& {Mendel}, J.~T. 2016,
  \mnras, 458, L34

\bibitem[{{Elvis} {et~al.}(1994){Elvis}, {Wilkes}, {McDowell}, {Green},
  {Bechtold}, {Willner}, {Oey}, {Polomski}, \& {Cutri}}]{EL94.1}
{Elvis}, M., {Wilkes}, B.~J., {McDowell}, J.~C., {Green}, R.~F., {Bechtold},
  J., {Willner}, S.~P., {Oey}, M.~S., {Polomski}, E., \& {Cutri}, R. 1994,
  \apjs, 95, 1

\bibitem[{{Fabian}(2012)}]{FA12.2}
{Fabian}, A.~C. 2012, \araa, 50, 455

\bibitem[{{Fanaroff} \& {Riley}(1974)}]{FA74.1}
{Fanaroff}, B.~L., \& {Riley}, J.~M. 1974, \mnras, 167, 31P

\bibitem[{{Gebhardt} {et~al.}(2000){Gebhardt}, {Bender}, {Bower}, {Dressler},
  {Faber}, {Filippenko}, {Green}, {Grillmair}, {Ho}, {Kormendy}, {Lauer},
  {Magorrian}, {Pinkney}, {Richstone}, \& {Tremaine}}]{GE00.1}
{Gebhardt}, K., {Bender}, R., {Bower}, G., {Dressler}, A., {Faber}, S.~M.,
  {Filippenko}, A.~V., {Green}, R., {Grillmair}, C., {Ho}, L.~C., {Kormendy},
  J., {Lauer}, T.~R., {Magorrian}, J., {Pinkney}, J., {Richstone}, D., \&
  {Tremaine}, S. 2000, \apjl, 539, L13

\bibitem[{{Gonz{\'a}lez Delgado} {et~al.}(2014){Gonz{\'a}lez Delgado},
  {P{\'e}rez}, {Cid Fernand es}, {Garc{\'\i}a-Benito}, {de Amorim},
  {S{\'a}nchez}, {Husemann}, {Cortijo-Ferrero}, {L{\'o}pez Fern{\'a}ndez},
  {S{\'a}nchez-Bl{\'a}zquez}, {Bekeraite}, {Walcher}, {Falc{\'o}n-Barroso},
  {Gallazzi}, {van de Ven}, {Alves}, {Bland -Hawthorn}, {Kennicutt}, {Kupko},
  {Lyubenova}, {Mast}, {Moll{\'a}}, {Marino}, {Quirrenbach}, {V{\'\i}lchez}, \&
  {Wisotzki}}]{GO14.1}
{Gonz{\'a}lez Delgado}, R.~M., {P{\'e}rez}, E., {Cid Fernand es}, R.,
  {Garc{\'\i}a-Benito}, R., {de Amorim}, A.~L., {S{\'a}nchez}, S.~F.,
  {Husemann}, B., {Cortijo-Ferrero}, C., {L{\'o}pez Fern{\'a}ndez}, R.,
  {S{\'a}nchez-Bl{\'a}zquez}, P., {Bekeraite}, S., {Walcher}, C.~J.,
  {Falc{\'o}n-Barroso}, J., {Gallazzi}, A., {van de Ven}, G., {Alves}, J.,
  {Bland -Hawthorn}, J., {Kennicutt}, R.~C., {Kupko}, D., {Lyubenova}, M.,
  {Mast}, D., {Moll{\'a}}, M., {Marino}, R.~A., {Quirrenbach}, A.,
  {V{\'\i}lchez}, J.~M., \& {Wisotzki}, L. 2014, \aap, 562, A47

\bibitem[{{Greene} \& {Ho}(2006)}]{GR06.1}
{Greene}, J.~E., \& {Ho}, L.~C. 2006, \apjl, 641, L21

\bibitem[{{Groves} {et~al.}(2006){Groves}, {Heckman}, \& {Kauffmann}}]{GR06.3}
{Groves}, B.~A., {Heckman}, T.~M., \& {Kauffmann}, G. 2006, \mnras, 371, 1559

\bibitem[{{Hardcastle}(2018)}]{HA18.1}
{Hardcastle}, M. 2018, Nature Astronomy, 2, 273

\bibitem[{{Heckman} \& {Best}(2014)}]{HE14.1}
{Heckman}, T.~M., \& {Best}, P.~N. 2014, \araa, 52, 589

\bibitem[{{Hickox} {et~al.}(2009){Hickox}, {Jones}, {Forman}, {Murray},
  {Kochanek}, {Eisenstein}, {Jannuzi}, {Dey}, {Brown}, \& {Stern}}]{HI09.1}
{Hickox}, R.~C., {Jones}, C., {Forman}, W.~R., {Murray}, S.~S., {Kochanek},
  C.~S., {Eisenstein}, D., {Jannuzi}, B.~T., {Dey}, A., {Brown}, M. J.~I., \&
  {Stern}, D. 2009, \apj, 696, 891

\bibitem[{{Hopkins} {et~al.}(2008){Hopkins}, {Hernquist}, {Cox}, \& {Kere{\v
  s}}}]{HO08.3}
{Hopkins}, P.~F., {Hernquist}, L., {Cox}, T.~J., \& {Kere{\v s}}, D. 2008,
  \apjs, 175, 356

\bibitem[{{Ibarra-Medel} {et~al.}(2019){Ibarra-Medel}, {Avila-Reese},
  {S{\'a}nchez}, {Gonz{\'a}lez-Samaniego}, \& {Rodr{\'\i}guez-Puebla}}]{IB19.1}
{Ibarra-Medel}, H.~J., {Avila-Reese}, V., {S{\'a}nchez}, S.~F.,
  {Gonz{\'a}lez-Samaniego}, A.~r., \& {Rodr{\'\i}guez-Puebla}, A. 2019, \mnras,
  483, 4525

\bibitem[{{Ichikawa} {et~al.}(2019){Ichikawa}, {Ricci}, {Ueda}, {Bauer},
  {Kawamuro}, {Koss}, {Oh}, {Rosario}, {Shimizu}, {Stalevski}, {Fuller},
  {Packham}, \& {Trakhtenbrot}}]{IC19.1}
{Ichikawa}, K., {Ricci}, C., {Ueda}, Y., {Bauer}, F.~E., {Kawamuro}, T.,
  {Koss}, M.~J., {Oh}, K., {Rosario}, D.~J., {Shimizu}, T.~T., {Stalevski}, M.,
  {Fuller}, L., {Packham}, C., \& {Trakhtenbrot}, B. 2019, \apj, 870, 31

\bibitem[{{Ishibashi} {et~al.}(2014){Ishibashi}, {Auger}, {Zhang}, \&
  {Fabian}}]{IS14.1}
{Ishibashi}, W., {Auger}, M.~W., {Zhang}, D., \& {Fabian}, A.~C. 2014, \mnras,
  443, 1339

\bibitem[{{Janssen} {et~al.}(2012){Janssen}, {R{\"o}ttgering}, {Best}, \&
  {Brinchmann}}]{JA12.1}
{Janssen}, R.~M.~J., {R{\"o}ttgering}, H.~J.~A., {Best}, P.~N., \&
  {Brinchmann}, J. 2012, \aap, 541, A62

\bibitem[{{Jarrett} {et~al.}(2011){Jarrett}, {Cohen}, {Masci}, {Wright},
  {Stern}, {Benford}, {Blain}, {Carey}, {Cutri}, {Eisenhardt}, {Lonsdale},
  {Mainzer}, {Marsh}, {Padgett}, {Petty}, {Ressler}, {Skrutskie}, {Stanford},
  {Surace}, {Tsai}, {Wheelock}, \& {Yan}}]{JA11.1}
{Jarrett}, T.~H., {Cohen}, M., {Masci}, F., {Wright}, E., {Stern}, D.,
  {Benford}, D., {Blain}, A., {Carey}, S., {Cutri}, R.~M., {Eisenhardt}, P.,
  {Lonsdale}, C., {Mainzer}, A., {Marsh}, K., {Padgett}, D., {Petty}, S.,
  {Ressler}, M., {Skrutskie}, M., {Stanford}, S., {Surace}, J., {Tsai}, C.~W.,
  {Wheelock}, S., \& {Yan}, D.~L. 2011, \apj, 735, 112

\bibitem[{{Ji} \& {Yan}(2020)}]{JI20.1}
{Ji}, X., \& {Yan}, R. 2020, arXiv e-prints, arXiv:2007.09159

\bibitem[{{Kalfountzou} {et~al.}(2012){Kalfountzou}, {Jarvis}, {Bonfield}, \&
  {Hardcastle}}]{KA12.1}
{Kalfountzou}, E., {Jarvis}, M.~J., {Bonfield}, D.~G., \& {Hardcastle}, M.~J.
  2012, \mnras, 427, 2401

\bibitem[{{Karouzos} {et~al.}(2014){Karouzos}, {Im}, {Trichas}, {Goto},
  {Malkan}, {Ruiz}, {Jeon}, {Kim}, {Lee}, \& {Kim}}]{KA14.1}
{Karouzos}, M., {Im}, M., {Trichas}, M., {Goto}, T., {Malkan}, M., {Ruiz}, A.,
  {Jeon}, Y., {Kim}, J.~H., {Lee}, H.~M., \& {Kim}, S.~J. 2014, \apj, 784, 137

\bibitem[{{Kauffmann} {et~al.}(2008){Kauffmann}, {Heckman}, \& {Best}}]{KA08.1}
{Kauffmann}, G., {Heckman}, T.~M., \& {Best}, P.~N. 2008, \mnras, 384, 953

\bibitem[{{Kewley} {et~al.}(2013){Kewley}, {Dopita}, {Leitherer}, {Dav{\'e}},
  {Yuan}, {Allen}, {Groves}, \& {Sutherland}}]{KE13.1}
{Kewley}, L.~J., {Dopita}, M.~A., {Leitherer}, C., {Dav{\'e}}, R., {Yuan}, T.,
  {Allen}, M., {Groves}, B., \& {Sutherland}, R. 2013, \apj, 774, 100

\bibitem[{{Kewley} {et~al.}(2001){Kewley}, {Dopita}, {Sutherland}, {Heisler},
  \& {Trevena}}]{KE01.2}
{Kewley}, L.~J., {Dopita}, M.~A., {Sutherland}, R.~S., {Heisler}, C.~A., \&
  {Trevena}, J. 2001, \apj, 556, 121

\bibitem[{{Kewley} {et~al.}(2006){Kewley}, {Groves}, {Kauffmann}, \&
  {Heckman}}]{KE06.1}
{Kewley}, L.~J., {Groves}, B., {Kauffmann}, G., \& {Heckman}, T. 2006, \mnras,
  372, 961

\bibitem[{{Kewley} {et~al.}(2019){Kewley}, {Nicholls}, \&
  {Sutherland}}]{KE19.1}
{Kewley}, L.~J., {Nicholls}, D.~C., \& {Sutherland}, R.~S. 2019, \araa, 57, 511

\bibitem[{{Law} {et~al.}(2016){Law}, {Cherinka}, {Yan}, {Andrews}, {Bershady},
  {Bizyaev}, {Blanc}, {Blanton}, {Bolton}, \& {Brownstein}}]{LA16.2}
{Law}, D.~R., {Cherinka}, B., {Yan}, R., {Andrews}, B.~H., {Bershady}, M.~A.,
  {Bizyaev}, D., {Blanc}, G.~A., {Blanton}, M.~R., {Bolton}, A.~S., \&
  {Brownstein}, J.~R. 2016, \aj, 152, 83

\bibitem[{{Law} {et~al.}(2015){Law}, {Yan}, {Bershady}, {Bundy}, {Cherinka},
  {Drory}, {MacDonald}, {S{\'a}nchez-Gallego}, {Wake}, {Weijmans}, {Blanton},
  {Klaene}, {Moran}, {Sanchez}, \& {Zhang}}]{LA15.1}
{Law}, D.~R., {Yan}, R., {Bershady}, M.~A., {Bundy}, K., {Cherinka}, B.,
  {Drory}, N., {MacDonald}, N., {S{\'a}nchez-Gallego}, J.~R., {Wake}, D.~A.,
  {Weijmans}, A.-M., {Blanton}, M.~R., {Klaene}, M.~A., {Moran}, S.~M.,
  {Sanchez}, S.~F., \& {Zhang}, K. 2015, \aj, 150, 19

\bibitem[{{Ledlow} \& {Owen}(1996)}]{LE96.2}
{Ledlow}, M.~J., \& {Owen}, F.~N. 1996, \aj, 112, 9

\bibitem[{{Leslie} {et~al.}(2016){Leslie}, {Kewley}, {Sanders}, \&
  {Lee}}]{LE16.1}
{Leslie}, S.~K., {Kewley}, L.~J., {Sanders}, D.~B., \& {Lee}, N. 2016, \mnras,
  455, L82

\bibitem[{{Mainzer} {et~al.}(2014){Mainzer}, {Bauer}, {Cutri}, {Grav},
  {Masiero}, {Beck}, {Clarkson}, {Conrow}, {Dailey}, {Eisenhardt}, {Fabinsky},
  {Fajardo-Acosta}, {Fowler}, {Gelino}, {Grillmair}, {Heinrichsen}, {Kendall},
  {Kirkpatrick}, {Liu}, {Masci}, {McCallon}, {Nugent}, {Papin}, {Rice},
  {Royer}, {Ryan}, {Sevilla}, {Sonnett}, {Stevenson}, {Thompson}, {Wheelock},
  {Wiemer}, {Wittman}, {Wright}, \& {Yan}}]{MA14.1}
{Mainzer}, A., {Bauer}, J., {Cutri}, R.~M., {Grav}, T., {Masiero}, J., {Beck},
  R., {Clarkson}, P., {Conrow}, T., {Dailey}, J., {Eisenhardt}, P., {Fabinsky},
  B., {Fajardo-Acosta}, S., {Fowler}, J., {Gelino}, C., {Grillmair}, C.,
  {Heinrichsen}, I., {Kendall}, M., {Kirkpatrick}, J.~D., {Liu}, F., {Masci},
  F., {McCallon}, H., {Nugent}, C.~R., {Papin}, M., {Rice}, E., {Royer}, D.,
  {Ryan}, T., {Sevilla}, P., {Sonnett}, S., {Stevenson}, R., {Thompson}, D.~B.,
  {Wheelock}, S., {Wiemer}, D., {Wittman}, M., {Wright}, E., \& {Yan}, L. 2014,
  \apj, 792, 30

\bibitem[{{Marchesi} {et~al.}(2018){Marchesi}, {Ajello}, {Marcotulli},
  {Comastri}, {Lanzuisi}, \& {Vignali}}]{MA18.1}
{Marchesi}, S., {Ajello}, M., {Marcotulli}, L., {Comastri}, A., {Lanzuisi}, G.,
  \& {Vignali}, C. 2018, \apj, 854, 49

\bibitem[{{Marconi} {et~al.}(2004){Marconi}, {Risaliti}, {Gilli}, {Hunt},
  {Maiolino}, \& {Salvati}}]{MA04.4}
{Marconi}, A., {Risaliti}, G., {Gilli}, R., {Hunt}, L.~K., {Maiolino}, R., \&
  {Salvati}, M. 2004, \mnras, 351, 169

\bibitem[{{Martig} {et~al.}(2009){Martig}, {Bournaud}, {Teyssier}, \&
  {Dekel}}]{MA09.1}
{Martig}, M., {Bournaud}, F., {Teyssier}, R., \& {Dekel}, A. 2009, \apj, 707,
  250

\bibitem[{{Matthews} {et~al.}(1964){Matthews}, {Morgan}, \& {Schmidt}}]{MA64.1}
{Matthews}, T.~A., {Morgan}, W.~W., \& {Schmidt}, M. 1964, \apj, 140, 35

\bibitem[{{McConnell} \& {Ma}(2013)}]{MC13.1}
{McConnell}, N.~J., \& {Ma}, C.-P. 2013, \apj, 764, 184

\bibitem[{{McPartland} {et~al.}(2019){McPartland}, {Sanders}, {Kewley}, \&
  {Leslie}}]{MC19.1}
{McPartland}, C., {Sanders}, D.~B., {Kewley}, L.~J., \& {Leslie}, S.~K. 2019,
  \mnras, 482, L129

\bibitem[{{Mendez} {et~al.}(2013){Mendez}, {Coil}, {Aird}, {Diamond-Stanic},
  {Moustakas}, {Blanton}, {Cool}, {Eisenstein}, {Wong}, \& {Zhu}}]{ME13.2}
{Mendez}, A.~J., {Coil}, A.~L., {Aird}, J., {Diamond-Stanic}, A.~M.,
  {Moustakas}, J., {Blanton}, M.~R., {Cool}, R.~J., {Eisenstein}, D.~J.,
  {Wong}, K.~C., \& {Zhu}, G. 2013, \apj, 770, 40

\bibitem[{{Merloni} \& {Heinz}(2007)}]{ME07.3}
{Merloni}, A., \& {Heinz}, S. 2007, \mnras, 381, 589

\bibitem[{{Morganti}(2010)}]{MO10.1}
{Morganti}, R. 2010, PASA, 27, 463

\bibitem[{{Morganti}(2017)}]{MO17.1}
---. 2017, Frontiers in Astronomy and Space Sciences, 4, 42

\bibitem[{{Mullaney} {et~al.}(2015){Mullaney}, {Alexander}, {Aird}, {Bernhard},
  {Daddi}, {Del Moro}, {Dickinson}, {Elbaz}, {Harrison}, \& {Juneau}}]{MU15.2}
{Mullaney}, J.~R., {Alexander}, D.~M., {Aird}, J., {Bernhard}, E., {Daddi}, E.,
  {Del Moro}, A., {Dickinson}, M., {Elbaz}, D., {Harrison}, C.~M., \& {Juneau},
  S. 2015, \mnras, 453, L83

\bibitem[{{Mullaney} {et~al.}(2013){Mullaney}, {Alexander}, {Fine}, {Goulding},
  {Harrison}, \& {Hickox}}]{MU13.3}
{Mullaney}, J.~R., {Alexander}, D.~M., {Fine}, S., {Goulding}, A.~D.,
  {Harrison}, C.~M., \& {Hickox}, R.~C. 2013, \mnras, 433, 622

\bibitem[{{Narayan} \& {Yi}(1995)}]{NA95.3}
{Narayan}, R., \& {Yi}, I. 1995, \apj, 452, 710

\bibitem[{{Nesvadba} {et~al.}(2010){Nesvadba}, {Boulanger}, {Salom{\'e}},
  {Guillard}, {Lehnert}, {Ogle}, {Appleton}, {Falgarone}, \& {Pineau Des
  Forets}}]{NE10.1}
{Nesvadba}, N.~P.~H., {Boulanger}, F., {Salom{\'e}}, P., {Guillard}, P.,
  {Lehnert}, M.~D., {Ogle}, P., {Appleton}, P., {Falgarone}, E., \& {Pineau Des
  Forets}, G. 2010, \aap, 521, A65

\bibitem[{{Nesvadba} {et~al.}(2006){Nesvadba}, {Lehnert}, {Eisenhauer},
  {Gilbert}, {Tecza}, \& {Abuter}}]{NE06.1}
{Nesvadba}, N.~P.~H., {Lehnert}, M.~D., {Eisenhauer}, F., {Gilbert}, A.,
  {Tecza}, M., \& {Abuter}, R. 2006, \apj, 650, 693

\bibitem[{{Noeske} {et~al.}(2007){Noeske}, {Weiner}, {Faber}, {Papovich},
  {Koo}, {Somerville}, {Bundy}, {Conselice}, {Newman}, {Schiminovich}, {Le
  Floc'h}, {Coil}, {Rieke}, {Lotz}, {Primack}, {Barmby}, {Cooper}, {Davis},
  {Ellis}, {Fazio}, {Guhathakurta}, {Huang}, {Kassin}, {Martin}, {Phillips},
  {Rich}, {Small}, {Willmer}, \& {Wilson}}]{NO07.1}
{Noeske}, K.~G., {Weiner}, B.~J., {Faber}, S.~M., {Papovich}, C., {Koo}, D.~C.,
  {Somerville}, R.~S., {Bundy}, K., {Conselice}, C.~J., {Newman}, J.~A.,
  {Schiminovich}, D., {Le Floc'h}, E., {Coil}, A.~L., {Rieke}, G.~H., {Lotz},
  J.~M., {Primack}, J.~R., {Barmby}, P., {Cooper}, M.~C., {Davis}, M., {Ellis},
  R.~S., {Fazio}, G.~G., {Guhathakurta}, P., {Huang}, J., {Kassin}, S.~A.,
  {Martin}, D.~C., {Phillips}, A.~C., {Rich}, R.~M., {Small}, T.~A., {Willmer},
  C.~N.~A., \& {Wilson}, G. 2007, \apjl, 660, L43

\bibitem[{{Oh} {et~al.}(2018){Oh}, {Koss}, {Markwardt}, {Schawinski},
  {Baumgartner}, {Barthelmy}, {Cenko}, {Gehrels}, {Mushotzky}, {Petulante},
  {Ricci}, {Lien}, \& {Trakhtenbrot}}]{OH18.1}
{Oh}, K., {Koss}, M., {Markwardt}, C.~B., {Schawinski}, K., {Baumgartner},
  W.~H., {Barthelmy}, S.~D., {Cenko}, S.~B., {Gehrels}, N., {Mushotzky}, R.,
  {Petulante}, A., {Ricci}, C., {Lien}, A., \& {Trakhtenbrot}, B. 2018, \apjs,
  235, 4

\bibitem[{{Oh} {et~al.}(2015){Oh}, {Yi}, {Schawinski}, {Koss}, {Trakhtenbrot},
  \& {Soto}}]{OH15.1}
{Oh}, K., {Yi}, S.~K., {Schawinski}, K., {Koss}, M., {Trakhtenbrot}, B., \&
  {Soto}, K. 2015, \apjs, 219, 1

\bibitem[{{Osterbrock}(1991)}]{OS91.1}
{Osterbrock}, D.~E. 1991, Reports of Progress in Physics, 54, 579

\bibitem[{{Panessa} {et~al.}(2015){Panessa}, {Tarchi}, {Castangia}, {Maiorano},
  {Bassani}, {Bicknell}, {Bazzano}, {Bird}, {Malizia}, \& {Ubertini}}]{PA15.1}
{Panessa}, F., {Tarchi}, A., {Castangia}, P., {Maiorano}, E., {Bassani}, L.,
  {Bicknell}, G., {Bazzano}, A., {Bird}, A.~J., {Malizia}, A., \& {Ubertini},
  P. 2015, \mnras, 447, 1289

\bibitem[{{Pawlik} \& {Schaye}(2009)}]{PA09.1}
{Pawlik}, A.~H., \& {Schaye}, J. 2009, \mnras, 396, L46

\bibitem[{{Pennell} {et~al.}(2017){Pennell}, {Runnoe}, \&
  {Brotherton}}]{PE17.1}
{Pennell}, A., {Runnoe}, J.~C., \& {Brotherton}, M.~S. 2017, \mnras, 468, 1433

\bibitem[{{Pitchford} {et~al.}(2016){Pitchford}, {Hatziminaoglou}, {Feltre},
  {Farrah}, {Clarke}, {Harris}, {Hurley}, {Oliver}, {Page}, \& {Wang}}]{PI16.1}
{Pitchford}, L.~K., {Hatziminaoglou}, E., {Feltre}, A., {Farrah}, D., {Clarke},
  C., {Harris}, K.~A., {Hurley}, P., {Oliver}, S., {Page}, M., \& {Wang}, L.
  2016, \mnras, 462, 4067

\bibitem[{{Ramos Almeida} {et~al.}(2012){Ramos Almeida}, {Bessiere},
  {Tadhunter}, {P{\'e}rez-Gonz{\'a}lez}, {Barro}, {Inskip}, {Morganti}, {Holt},
  \& {Dicken}}]{RA12.1}
{Ramos Almeida}, C., {Bessiere}, P.~S., {Tadhunter}, C.~N.,
  {P{\'e}rez-Gonz{\'a}lez}, P.~G., {Barro}, G., {Inskip}, K.~J., {Morganti},
  R., {Holt}, J., \& {Dicken}, D. 2012, \mnras, 419, 687

\bibitem[{{Ramos Almeida} {et~al.}(2011){Ramos Almeida}, {Tadhunter}, {Inskip},
  {Morganti}, {Holt}, \& {Dicken}}]{RA11.1}
{Ramos Almeida}, C., {Tadhunter}, C.~N., {Inskip}, K.~J., {Morganti}, R.,
  {Holt}, J., \& {Dicken}, D. 2011, \mnras, 410, 1550

\bibitem[{{Rembold} {et~al.}(2017){Rembold}, {Shimoia}, {Storchi-Bergmann},
  {Riffel}, {Riffel}, {Mallmann}, {do Nascimento}, {Moreira}, {Ilha},
  {Machado}, {Cirolini}, {da Costa}, {Maia}, {Santiago}, {Schneider},
  {Wylezalek}, {Bizyaev}, {Pan}, \& {M{\"u}ller-S{\'a}nchez}}]{RE17.1}
{Rembold}, S.~B., {Shimoia}, J.~S., {Storchi-Bergmann}, T., {Riffel}, R.,
  {Riffel}, R.~A., {Mallmann}, N.~D., {do Nascimento}, J.~C., {Moreira}, T.~N.,
  {Ilha}, G.~S., {Machado}, A.~D., {Cirolini}, R., {da Costa}, L.~N., {Maia},
  M.~A.~G., {Santiago}, B.~X., {Schneider}, D.~P., {Wylezalek}, D., {Bizyaev},
  D., {Pan}, K., \& {M{\"u}ller-S{\'a}nchez}, F. 2017, \mnras, 472, 4382

\bibitem[{{Ricci} {et~al.}(2017){Ricci}, {Trakhtenbrot}, {Koss}, {Ueda}, {Del
  Vecchio}, {Treister}, {Schawinski}, {Paltani}, {Oh}, {Lamperti}, {Berney},
  {Gandhi}, {Ichikawa}, {Bauer}, {Ho}, {Asmus}, {Beckmann}, {Soldi},
  {Balokovi{\'c}}, {Gehrels}, \& {Markwardt}}]{RI17.2}
{Ricci}, C., {Trakhtenbrot}, B., {Koss}, M.~J., {Ueda}, Y., {Del Vecchio}, I.,
  {Treister}, E., {Schawinski}, K., {Paltani}, S., {Oh}, K., {Lamperti}, I.,
  {Berney}, S., {Gandhi}, P., {Ichikawa}, K., {Bauer}, F.~E., {Ho}, L.~C.,
  {Asmus}, D., {Beckmann}, V., {Soldi}, S., {Balokovi{\'c}}, M., {Gehrels}, N.,
  \& {Markwardt}, C.~B. 2017, \apjs, 233, 17

\bibitem[{{Rich} {et~al.}(2011){Rich}, {Kewley}, \& {Dopita}}]{RI11.1}
{Rich}, J.~A., {Kewley}, L.~J., \& {Dopita}, M.~A. 2011, \apj, 734, 87

\bibitem[{{Rosario} {et~al.}(2013){Rosario}, {Trakhtenbrot}, {Lutz}, {Netzer},
  {Trump}, {Silverman}, {Schramm}, {Lusso}, {Berta}, \& {Bongiorno}}]{RO13.1}
{Rosario}, D.~J., {Trakhtenbrot}, B., {Lutz}, D., {Netzer}, H., {Trump}, J.~R.,
  {Silverman}, J.~D., {Schramm}, M., {Lusso}, E., {Berta}, S., \& {Bongiorno},
  A. 2013, \aap, 560, A72

\bibitem[{{S{\'a}nchez}(2020)}]{SA20.1}
{S{\'a}nchez}, S.~F. 2020, \araa, 58, annurev

\bibitem[{{S{\'a}nchez} {et~al.}(2018){S{\'a}nchez}, {Avila-Reese},
  {Hernandez-Toledo}, {Cortes-Su{\'a}rez}, {Rodr{\'{\i}}guez-Puebla},
  {Ibarra-Medel}, {Cano-D{\'{\i}}az}, {Barrera-Ballesteros}, {Negrete},
  {Calette}, {de Lorenzo-C{\'a}ceres}, {Ortega-Minakata}, {Aquino},
  {Valenzuela}, {Clemente}, {Storchi-Bergmann}, {Riffel}, {Schimoia}, {Riffel},
  {Rembold}, {Brownstein}, {Pan}, {Yates}, {Mallmann}, \& {Bitsakis}}]{SA18.2}
{S{\'a}nchez}, S.~F., {Avila-Reese}, V., {Hernandez-Toledo}, H.,
  {Cortes-Su{\'a}rez}, E., {Rodr{\'{\i}}guez-Puebla}, A., {Ibarra-Medel}, H.,
  {Cano-D{\'{\i}}az}, M., {Barrera-Ballesteros}, J.~K., {Negrete}, C.~A.,
  {Calette}, A.~R., {de Lorenzo-C{\'a}ceres}, A., {Ortega-Minakata}, R.~A.,
  {Aquino}, E., {Valenzuela}, O., {Clemente}, J.~C., {Storchi-Bergmann}, T.,
  {Riffel}, R., {Schimoia}, J., {Riffel}, R.~A., {Rembold}, S.~B.,
  {Brownstein}, J.~R., {Pan}, K., {Yates}, R., {Mallmann}, N., \& {Bitsakis},
  T. 2018, RMxAA, 54, 217

\bibitem[{{S{\'a}nchez} {et~al.}(2019){S{\'a}nchez}, {Avila-Reese},
  {Rodr{\'\i}guez-Puebla}, {Ibarra-Medel}, {Calette}, {Bershady},
  {Hern{\'a}ndez-Toledo}, {Pan}, \& {Bizyaev}}]{SA19.1}
{S{\'a}nchez}, S.~F., {Avila-Reese}, V., {Rodr{\'\i}guez-Puebla}, A.,
  {Ibarra-Medel}, H., {Calette}, R., {Bershady}, M., {Hern{\'a}ndez-Toledo},
  H., {Pan}, K., \& {Bizyaev}, D. 2019, \mnras, 482, 1557

\bibitem[{{S{\'a}nchez} {et~al.}(2016){S{\'a}nchez}, {P{\'e}rez},
  {S{\'a}nchez-Bl{\'a}zquez}, {Garc{\'{\i}}a-Benito}, {Ibarra-Mede},
  {Gonz{\'a}lez}, {Rosales-Ortega}, {S{\'a}nchez-Menguiano}, {Ascasibar},
  {Bitsakis}, {Law}, {Cano-D{\'{\i}}az}, {L{\'o}pez-Cob{\'a}}, {Marino}, {Gil
  de Paz}, {L{\'o}pez-S{\'a}nchez}, {Barrera-Ballesteros}, {Galbany}, {Mast},
  {Abril-Melgarejo}, \& {Roman-Lopes}}]{SA16.1}
{S{\'a}nchez}, S.~F., {P{\'e}rez}, E., {S{\'a}nchez-Bl{\'a}zquez}, P.,
  {Garc{\'{\i}}a-Benito}, R., {Ibarra-Mede}, H.~J., {Gonz{\'a}lez}, J.~J.,
  {Rosales-Ortega}, F.~F., {S{\'a}nchez-Menguiano}, L., {Ascasibar}, Y.,
  {Bitsakis}, T., {Law}, D., {Cano-D{\'{\i}}az}, M., {L{\'o}pez-Cob{\'a}}, C.,
  {Marino}, R.~A., {Gil de Paz}, A., {L{\'o}pez-S{\'a}nchez}, A.~R.,
  {Barrera-Ballesteros}, J., {Galbany}, L., {Mast}, D., {Abril-Melgarejo}, V.,
  \& {Roman-Lopes}, A. 2016, RMxAA, 52, 171

\bibitem[{{Schawinski} {et~al.}(2010){Schawinski}, {Urry}, {Virani}, {Coppi},
  {Bamford}, {Treister}, {Lintott}, {Sarzi}, {Keel}, {Kaviraj}, {Cardamone},
  {Masters}, {Ross}, {Andreescu}, {Murray}, {Nichol}, {Raddick}, {Slosar},
  {Szalay}, {Thomas}, \& {Vandenberg}}]{SC10.1}
{Schawinski}, K., {Urry}, C.~M., {Virani}, S., {Coppi}, P., {Bamford}, S.~P.,
  {Treister}, E., {Lintott}, C.~J., {Sarzi}, M., {Keel}, W.~C., {Kaviraj}, S.,
  {Cardamone}, C.~N., {Masters}, K.~L., {Ross}, N.~P., {Andreescu}, D.,
  {Murray}, P., {Nichol}, R.~C., {Raddick}, M.~J., {Slosar}, A., {Szalay},
  A.~S., {Thomas}, D., \& {Vandenberg}, J. 2010, \apj, 711, 284

\bibitem[{{Shakura} \& {Sunyaev}(1973)}]{SH73.1}
{Shakura}, N.~I., \& {Sunyaev}, R.~A. 1973, \aap, 500, 33

\bibitem[{{Shimizu} {et~al.}(2015){Shimizu}, {Mushotzky}, {Mel{\'e}ndez},
  {Koss}, \& {Rosario}}]{SH15.2}
{Shimizu}, T.~T., {Mushotzky}, R.~F., {Mel{\'e}ndez}, M., {Koss}, M., \&
  {Rosario}, D.~J. 2015, \mnras, 452, 1841

\bibitem[{{Shimizu} {et~al.}(2017){Shimizu}, {Mushotzky}, {Mel{\'e}ndez},
  {Koss}, {Barger}, \& {Cowie}}]{SH17.1}
{Shimizu}, T.~T., {Mushotzky}, R.~F., {Mel{\'e}ndez}, M., {Koss}, M.~J.,
  {Barger}, A.~J., \& {Cowie}, L.~L. 2017, \mnras, 466, 3161

\bibitem[{{Silk} \& {Mamon}(2012)}]{SI12.1}
{Silk}, J., \& {Mamon}, G.~A. 2012, Research in Astronomy and Astrophysics, 12,
  917

\bibitem[{{Silk} \& {Rees}(1998)}]{SI98.1}
{Silk}, J., \& {Rees}, M.~J. 1998, \aap, 331, L1

\bibitem[{{Smol{\v{c}}i{\'c}}(2009)}]{SM09.3}
{Smol{\v{c}}i{\'c}}, V. 2009, \apjl, 699, L43

\bibitem[{{Speagle} {et~al.}(2014){Speagle}, {Steinhardt}, {Capak}, \&
  {Silverman}}]{SP14.1}
{Speagle}, J.~S., {Steinhardt}, C.~L., {Capak}, P.~L., \& {Silverman}, J.~D.
  2014, \apjs, 214, 15

\bibitem[{{Stanley} {et~al.}(2017){Stanley}, {Alexander}, {Harrison},
  {Rosario}, {Wang}, {Aird}, {Bourne}, {Dunne}, {Dye}, \& {Eales}}]{ST17.1}
{Stanley}, F., {Alexander}, D.~M., {Harrison}, C.~M., {Rosario}, D.~J., {Wang},
  L., {Aird}, J.~A., {Bourne}, N., {Dunne}, L., {Dye}, S., \& {Eales}, S. 2017,
  \mnras, 472, 2221

\bibitem[{{Stern}(2015)}]{ST15.1}
{Stern}, D. 2015, \apj, 807, 129

\bibitem[{{Stern} {et~al.}(2012){Stern}, {Assef}, {Benford}, {Blain}, {Cutri},
  {Dey}, {Eisenhardt}, {Griffith}, {Jarrett}, {Lake}, {Masci}, {Petty},
  {Stanford}, {Tsai}, {Wright}, {Yan}, {Harrison}, \& {Madsen}}]{ST12.2}
{Stern}, D., {Assef}, R.~J., {Benford}, D.~J., {Blain}, A., {Cutri}, R., {Dey},
  A., {Eisenhardt}, P., {Griffith}, R.~L., {Jarrett}, T.~H., {Lake}, S.,
  {Masci}, F., {Petty}, S., {Stanford}, S.~A., {Tsai}, C.-W., {Wright}, E.~L.,
  {Yan}, L., {Harrison}, F., \& {Madsen}, K. 2012, \apj, 753, 30

\bibitem[{{Storchi-Bergmann} \& {Schnorr-M{\"u}ller}(2019)}]{ST19.1}
{Storchi-Bergmann}, T., \& {Schnorr-M{\"u}ller}, A. 2019, Nature Astronomy, 3,
  48

\bibitem[{{Strateva} {et~al.}(2001){Strateva}, {Ivezi{\'c}}, {Knapp},
  {Narayanan}, {Strauss}, {Gunn}, {Lupton}, {Schlegel}, {Bahcall}, {Brinkmann},
  {Brunner}, {Budav{\'a}ri}, {Csabai}, {Castander}, {Doi}, {Fukugita}, {Gy{\H
  o}ry}, {Hamabe}, {Hennessy}, {Ichikawa}, {Kunszt}, {Lamb}, {McKay},
  {Okamura}, {Racusin}, {Sekiguchi}, {Schneider}, {Shimasaku}, \&
  {York}}]{ST01.2}
{Strateva}, I., {Ivezi{\'c}}, {\v Z}., {Knapp}, G.~R., {Narayanan}, V.~K.,
  {Strauss}, M.~A., {Gunn}, J.~E., {Lupton}, R.~H., {Schlegel}, D., {Bahcall},
  N.~A., {Brinkmann}, J., {Brunner}, R.~J., {Budav{\'a}ri}, T., {Csabai}, I.,
  {Castander}, F.~J., {Doi}, M., {Fukugita}, M., {Gy{\H o}ry}, Z., {Hamabe},
  M., {Hennessy}, G., {Ichikawa}, T., {Kunszt}, P.~Z., {Lamb}, D.~Q., {McKay},
  T.~A., {Okamura}, S., {Racusin}, J., {Sekiguchi}, M., {Schneider}, D.~P.,
  {Shimasaku}, K., \& {York}, D. 2001, \aj, 122, 1861

\bibitem[{{Tadhunter} {et~al.}(1998){Tadhunter}, {Morganti}, {Robinson},
  {Dickson}, {Villar-Martin}, \& {Fosbury}}]{TA98.1}
{Tadhunter}, C.~N., {Morganti}, R., {Robinson}, A., {Dickson}, R.,
  {Villar-Martin}, M., \& {Fosbury}, R.~A.~E. 1998, \mnras, 298, 1035

\bibitem[{{Tasse} {et~al.}(2008){Tasse}, {Best}, {R{\"o}ttgering}, \& {Le
  Borgne}}]{TA08.1}
{Tasse}, C., {Best}, P.~N., {R{\"o}ttgering}, H., \& {Le Borgne}, D. 2008,
  \aap, 490, 893

\bibitem[{{van Zee} {et~al.}(1998){van Zee}, {Salzer}, \& {Haynes}}]{VA98.2}
{van Zee}, L., {Salzer}, J.~J., \& {Haynes}, M.~P. 1998, \apjl, 497, L1

\bibitem[{{Vasudevan} \& {Fabian}(2007)}]{VA07.1}
{Vasudevan}, R.~V., \& {Fabian}, A.~C. 2007, \mnras, 381, 1235

\bibitem[{{V{\'e}ron-Cetty} \& {V{\'e}ron}(2001)}]{VE01.2}
{V{\'e}ron-Cetty}, M.~P., \& {V{\'e}ron}, P. 2001, \aap, 375, 791

\bibitem[{{Wagner} \& {Bicknell}(2011)}]{WA11.4}
{Wagner}, A.~Y., \& {Bicknell}, G.~V. 2011, \apj, 728, 29

\bibitem[{{Wake} {et~al.}(2017){Wake}, {Bundy}, {Diamond-Stanic}, {Yan},
  {Blanton}, {Bershady}, {S{\'a}nchez-Gallego}, {Drory}, {Jones}, {Kauffmann},
  {Law}, {Li}, {MacDonald}, {Masters}, {Thomas}, {Tinker}, {Weijmans}, \&
  {Brownstein}}]{WA17.1}
{Wake}, D.~A., {Bundy}, K., {Diamond-Stanic}, A.~M., {Yan}, R., {Blanton},
  M.~R., {Bershady}, M.~A., {S{\'a}nchez-Gallego}, J.~R., {Drory}, N., {Jones},
  A., {Kauffmann}, G., {Law}, D.~R., {Li}, C., {MacDonald}, N., {Masters}, K.,
  {Thomas}, D., {Tinker}, J., {Weijmans}, A.-M., \& {Brownstein}, J.~R. 2017,
  \aj, 154, 86

\bibitem[{{White} {et~al.}(2000){White}, {Becker}, {Gregg},
  {Laurent-Muehleisen}, {Brotherton}, {Impey}, {Petry}, {Foltz}, {Chaffee},
  {Richards}, {Oegerle}, {Helfand}, {McMahon}, \& {Cabanela}}]{WH00.2}
{White}, R.~L., {Becker}, R.~H., {Gregg}, M.~D., {Laurent-Muehleisen}, S.~A.,
  {Brotherton}, M.~S., {Impey}, C.~D., {Petry}, C.~E., {Foltz}, C.~B.,
  {Chaffee}, F.~H., {Richards}, G.~T., {Oegerle}, W.~R., {Helfand}, D.~J.,
  {McMahon}, R.~G., \& {Cabanela}, J.~E. 2000, \apjs, 126, 133

\bibitem[{{Willett} {et~al.}(2013){Willett}, {Lintott}, {Bamford}, {Masters},
  {Simmons}, {Casteels}, {Edmondson}, {Fortson}, {Kaviraj}, {Keel}, {Melvin},
  {Nichol}, {Raddick}, {Schawinski}, {Simpson}, {Skibba}, {Smith}, \&
  {Thomas}}]{WI13.1}
{Willett}, K.~W., {Lintott}, C.~J., {Bamford}, S.~P., {Masters}, K.~L.,
  {Simmons}, B.~D., {Casteels}, K. R.~V., {Edmondson}, E.~M., {Fortson}, L.~F.,
  {Kaviraj}, S., {Keel}, W.~C., {Melvin}, T., {Nichol}, R.~C., {Raddick},
  M.~J., {Schawinski}, K., {Simpson}, R.~J., {Skibba}, R.~A., {Smith}, A.~M.,
  \& {Thomas}, D. 2013, \mnras, 435, 2835

\bibitem[{{Wright} {et~al.}(2010){Wright}, {Eisenhardt}, {Mainzer}, {Ressler},
  {Cutri}, {Jarrett}, {Kirkpatrick}, {Padgett}, {McMillan}, {Skrutskie},
  {Stanford}, {Cohen}, {Walker}, {Mather}, {Leisawitz}, {Gautier}, {McLean},
  {Benford}, {Lonsdale}, {Blain}, {Mendez}, {Irace}, {Duval}, {Liu}, {Royer},
  {Heinrichsen}, {Howard}, {Shannon}, {Kendall}, {Walsh}, {Larsen}, {Cardon},
  {Schick}, {Schwalm}, {Abid}, {Fabinsky}, {Naes}, \& {Tsai}}]{WR10.1}
{Wright}, E.~L., {Eisenhardt}, P.~R.~M., {Mainzer}, A.~K., {Ressler}, M.~E.,
  {Cutri}, R.~M., {Jarrett}, T., {Kirkpatrick}, J.~D., {Padgett}, D.,
  {McMillan}, R.~S., {Skrutskie}, M., {Stanford}, S.~A., {Cohen}, M., {Walker},
  R.~G., {Mather}, J.~C., {Leisawitz}, D., {Gautier}, III, T.~N., {McLean}, I.,
  {Benford}, D., {Lonsdale}, C.~J., {Blain}, A., {Mendez}, B., {Irace}, W.~R.,
  {Duval}, V., {Liu}, F., {Royer}, D., {Heinrichsen}, I., {Howard}, J.,
  {Shannon}, M., {Kendall}, M., {Walsh}, A.~L., {Larsen}, M., {Cardon}, J.~G.,
  {Schick}, S., {Schwalm}, M., {Abid}, M., {Fabinsky}, B., {Naes}, L., \&
  {Tsai}, C.-W. 2010, \aj, 140, 1868

\bibitem[{{Wylezalek} \& {Morganti}(2018)}]{WY18.2}
{Wylezalek}, D., \& {Morganti}, R. 2018, Nature Astronomy, 2, 181

\bibitem[{{Wylezalek} {et~al.}(2018){Wylezalek}, {Zakamska}, {Greene},
  {Riffel}, {Drory}, {Andrews}, {Merloni}, \& {Thomas}}]{WY18.1}
{Wylezalek}, D., {Zakamska}, N.~L., {Greene}, J.~E., {Riffel}, R.~A., {Drory},
  N., {Andrews}, B.~H., {Merloni}, A., \& {Thomas}, D. 2018, \mnras, 474, 1499

\bibitem[{{Yan} \& {Blanton}(2012)}]{YA12.1}
{Yan}, R., \& {Blanton}, M.~R. 2012, \apj, 747, 61

\bibitem[{{Yan} {et~al.}(2016){Yan}, {Bundy}, {Law}, {Bershady}, {Andrews},
  {Cherinka}, {Diamond-Stanic}, {Drory}, {MacDonald}, \&
  {S{\'a}nchez-Gallego}}]{YA16.1}
{Yan}, R., {Bundy}, K., {Law}, D.~R., {Bershady}, M.~A., {Andrews}, B.,
  {Cherinka}, B., {Diamond-Stanic}, A.~M., {Drory}, N., {MacDonald}, N., \&
  {S{\'a}nchez-Gallego}, J.~R. 2016, \aj, 152, 197

\bibitem[{{Young} {et~al.}(2014){Young}, {Eracleous}, {Shemmer}, {Netzer},
  {Gronwall}, {Lutz}, {Ciardullo}, \& {Sturm}}]{YO14.1}
{Young}, J.~E., {Eracleous}, M., {Shemmer}, O., {Netzer}, H., {Gronwall}, C.,
  {Lutz}, D., {Ciardullo}, R., \& {Sturm}, E. 2014, \mnras, 438, 217

\bibitem[{{Zheng} {et~al.}(2017){Zheng}, {Wang}, {Ge}, {Mao}, {Li}, {Li}, {Mo},
  {Goddard}, {Bundy}, {Li}, {Nair}, {Lin}, {Long}, {Riffel}, {Thomas},
  {Masters}, {Bizyaev}, {Brownstein}, {Zhang}, {Law}, {Drory}, {Roman Lopes},
  \& {Malanushenko}}]{ZH17.1}
{Zheng}, Z., {Wang}, H., {Ge}, J., {Mao}, S., {Li}, C., {Li}, R., {Mo}, H.,
  {Goddard}, D., {Bundy}, K., {Li}, H., {Nair}, P., {Lin}, L., {Long}, R.~J.,
  {Riffel}, R., {Thomas}, D., {Masters}, K., {Bizyaev}, D., {Brownstein},
  J.~R., {Zhang}, K., {Law}, D.~R., {Drory}, N., {Roman Lopes}, A., \&
  {Malanushenko}, O. 2017, \mnras, 465, 4572

\bibitem[{{Zinn} {et~al.}(2013){Zinn}, {Middelberg}, {Norris}, \&
  {Dettmar}}]{ZI13.1}
{Zinn}, P.-C., {Middelberg}, E., {Norris}, R.~P., \& {Dettmar}, R.-J. 2013,
  \apj, 774, 66

\end{thebibliography}

\end{document}